\documentclass[12pt]{article}
\pdfoutput=1

\usepackage{graphics,amssymb,float}
\usepackage{graphicx}% Include figure files
\usepackage{subfigure}
\usepackage{color}
\usepackage{xspace}
\usepackage{placeins}
\usepackage{rotating}% Include figure files
\usepackage{dcolumn}% Align table columns on decimal point
\usepackage{bm}% bold math
\usepackage{cite}
\usepackage{amsmath}
\usepackage{indentfirst} % activate indent in the first paragraph
\usepackage{epstopdf}
\usepackage{tikz}
\usepackage{longtable,multirow}
\usetikzlibrary{shapes,arrows}
\usepackage{appendix}

\def\smodelsnn  {{\sc SModelS}}
\def\smodels  {{\sc SModelS}\,v1.0.1}

\def\eg{{\it e.g.}}
\def\ie{{\it i.e.}}

\newcommand{\br}{\mbox{\ensuremath{\mathcal{B}}}}
\newcommand{\sigmaXBF}{\mbox{\ensuremath{\sigma\times\mathcal{B}}}\xspace}

\newcommand{\stau}{\tilde{\tau}_1^-}
\newcommand{\snu}{\tilde{\nu}}
\newcommand{\neu}{\widetilde{\chi}^0}
\newcommand{\cha}{\widetilde{\chi}^\pm}
\newcommand{\chap}{\widetilde{\chi}^+}
\newcommand{\cham}{\widetilde{\chi}^-}

\newcommand{\der}{{\rm d}}

\renewcommand{\ell}{{l}}

\def\lsim{\mathrel{\raise.3ex\hbox{$<$\kern-.75em\lower1ex\hbox{$\sim$}}}}
\def\gsim{\mathrel{\raise.3ex\hbox{$>$\kern-.75em\lower1ex\hbox{$\sim$}}}}
\def\ifmath#1{\relax\ifmmode #1\else $#1$\fi}

\begin{document}\setcounter{page}{0}\thispagestyle{empty}
%%%%%%%%%%%%%%%%%%%%%%%%%%%%%%%%%%%%%%%%%%%%%%%%%%%%%%%%%%%%
\begin{center}

\vspace*{-1cm}
\begin{flushright}
LPSC15065\\
HEPHY-PUB 948/15
\end{flushright}

\vspace*{2cm}

{\Large\bf Constraints on sneutrino dark matter from LHC Run~1} 

\vspace*{1.4cm}

\renewcommand{\thefootnote}{\fnsymbol{footnote}}

{\large 
Chiara~Arina$^{1,2}$\footnote[1]{Email: carina@uva.nl}, 
Maria~Eugenia~Cabrera~Catalan$^{2,3,4}$\footnote[2]{Email: mcabrera@if.usp.br},
Sabine~Kraml$^{5}$\footnote[3]{Email: sabine.kraml@lpsc.in2p3.fr},
Suchita~Kulkarni$^{5,6}$\footnote[4]{Email: suchita.kulkarni@oeaw.ac.at},
Ursula~Laa$^{5,6}$\footnote[5]{Email: ursula.laa@lpsc.in2p3.fr}\\[3mm]
} 

\renewcommand{\thefootnote}{\arabic{footnote}}

\vspace*{1cm} 
{\normalsize \it 
$^1\,$Institute d'Astrophysique de Paris, 98bis Boulevard Arago, F-75014 Paris, France\\[2mm]
$^2\,$GRAPPA Institute, University of Amsterdam, Science Park 904, 1090 GL Amsterdam, Netherlands\\[2mm]
$^3\,$Instituto de F\'{i}sica, Universidade de S\~{a}o Paulo, S\~{a}o Paulo, Brazil\\[2mm]
$^4\,$Instituto de F\'{i}sica Te\'orica, IFT-UAM/CSIC, U.A.M. Cantoblanco, 28049 Madrid, Spain\\[2mm]
$^5\,$Laboratoire de Physique Subatomique et de Cosmologie, Universit\'e Grenoble-Alpes,
CNRS/IN2P3, 53 Avenue des Martyrs, F-38026 Grenoble, France\\[2mm]
$^6\,$Institut f\"ur Hochenergiephysik,  \"Osterreichische Akademie der Wissenschaften,\\ Nikolsdorfer Gasse 18, 1050 Wien, Austria\\[2mm]
}

\vspace{1cm}

\begin{abstract}
A mostly right-handed sneutrino as the lightest supersymmetric particle (LSP) is an interesting dark matter candidate, leading to  LHC signatures which can be quite distinct from those of the conventional neutralino LSP. Using \smodels\ for testing the model against the limits published by ATLAS and CMS in the context of  so-called Simplified Model Spectra (SMS), we investigate to what extent the supersymmetry searches at Run~1 of the LHC constrain the sneutrino-LSP scenario. 
Moreover, we discuss the most relevant topologies for which no SMS results are provided by the experimental collaborations but which would allow to put more stringent constraints on sneutrino LSPs. These include, for instance, the mono-lepton signature which should be particularly interesting to consider at Run~2 of the LHC.
\end{abstract}

\end{center}

\clearpage

%%%%%%%%%%%%%%%%%%%%%%%%%%%%%%%%%%%%%%%%%%%%%%%%%%%%%%%%%%%%
\section{Introduction}\label{sec:intro}
%%%%%%%%%%%%%%%%%%%%%%%%%%%%%%%%%%%%%%%%%%%%%%%%%%%%%%%%%%%%

Before the start of data taking at the LHC, the common perception was that
supersymmetry (SUSY), if it has anything to do with stabilizing the
electroweak scale, would be discovered quickly, while Higgs physics would need
to wait for rather high statistics. In reality, quite the opposite has
happened: a Higgs boson has been found \cite{atlas:2012gk,cms:2012gu}, but
there is still no sign of SUSY---or of any new physics beyond the Standard
Model (BSM) whatsoever.

Indeed, the searches at Run~1 of the LHC at centre-of-mass energies of 7--8~TeV %in 2010--2012, 
have pushed the mass limits of SUSY particles quite high already, well above 1~TeV for 1st/2nd generation squarks 
and gluinos~\cite{atlas:susy:twiki,cms:susy:twiki}. 
Scenarios with high-scale \cite{Djouadi:2013vqa}, 
split~\cite{Giudice:2004tc,Bhattacharyya:2012ct,Benakli:2013msa} 
or at least spread~\cite{Hall:2011jd} SUSY 
are thus becoming increasingly popular in the literature. 
It should be kept in mind, however, that the current LHC limits sensitively depend on the presence of particular decay modes, 
and are considerably weakened in case of compressed \cite{Dreiner:2012gx} or stealth \cite{Fan:2011yu} spectra. 
Besides, the squark/gluino mass limits vanish completely in case the neutralino LSP is heavier than about 600~GeV. 

It should also be kept in mind that the SUSY mass limits depend sensitively on the nature of the LSP. 
Most experimental analyses indeed assume that the LSP is the lightest neutralino of the Minimal 
Supersymmetric Standard Model (MSSM). 
A particularly interesting alternative, and the subject of this paper, 
is a mainly right-handed (RH) mixed sneutrino in the MSSM augmented by a RH neutrino superfield~\cite{Borzumati:2000mc,Arkani-Hamed:2000bq}. 
This case is well motivated by two basic problems: the origin of neutrino masses and the nature of dark matter (DM). 
Its LHC signatures can be quite distinct from those of the conventional neutralino LSP. 

The left-handed (LH) sneutrino of the MSSM is excluded as the LSP and as a DM candidate because it has a non-zero hypercharge: its couplings to the $Z$ boson makes it annihilate too efficiently in the early Universe, and hence its final relic abundance is lower than the value $\Omega_{\rm DM} h^2$ measured by the WMAP and Planck satellites~\cite{Hinshaw:2012aka,Ade:2013zuv}. Very stringent limits come moreover from direct DM detection experiments: the $\tilde{\nu}_L$ scattering off nuclei is mediated by $t$-channel $Z$ boson exchange, giving a spin-independent (SI) cross section of order $10^{-39} {\rm cm^2}$ --- a value excluded already a decade ago for DM particles heavier than 10~GeV. 
A light $\tilde{\nu}_L$ with mass below $m_Z/2$ is also excluded by the $Z$ invisible width. 

The picture changes dramatically if we include in the MSSM a RH neutrino superfield (MSSM+RN from here on), which gives rise to Dirac neutrino masses. 
Besides the RH neutrino, the superfield also contains a scalar field, the RH sneutrino $\tilde{N}$ (strictly speaking this is a right-chiral field, but we use the RH notation for simplicity). This field, if at TeV scale, can mix with the LH partner $\tilde{\nu}_L$ and yield a mostly RH sneutrino LSP as a viable thermal DM candidate~\cite{Borzumati:2000mc,Arkani-Hamed:2000bq}.\footnote{Pure right-handed sterile sneutrinos can also be viable (non-thermal, depending on the model) DM candidates, as discussed {\it e.g.} in~\cite{Asaka:2005cn,Deppisch:2008bp,Cerdeno:2009dv,Khalil:2011tb,Choi:2012ap}.}

The phenomenology of this model was investigated in detail in~\cite{Arkani-Hamed:2000bq,Belanger:2010cd,Dumont:2012ee}. 
Indirect detection and cosmology were discussed in~\cite{Hooper:2004dc,Arina:2007tm,Choi:2012ap}, 
and LHC signatures in~\cite{Thomas:2007bu,Belanger:2011ny,Arina:2013zca} (see also \cite{BhupalDev:2012ru,Guo:2013asa,Harland-Lang:2013wxa} for related LHC studies). 
Reference~\cite{Arina:2013zca} 
in fact gave an update of the status of the sneutrino as DM after the Higgs mass measurements, 
by exploring the SUSY parameter space with the soft breaking terms fixed at the grand unification (GUT) scale, 
and assessing also the impact of the most recent exclusion bound for DM direct searches from LUX~\cite{Akerib:2013tjd}. 

In this paper, we extend the work of \cite{Arina:2013zca} by investigating to what extent the results from 
SUSY searches at Run~1 of the LHC, published in terms of so-called Simplified Model Spectra 
(SMS) limits,\footnote{Simplified Models are effective-Lagrangian descriptions involving only a small number of new particles. They were designed as a useful tool for the characterization of new physics, see \eg\ \cite{Alwall:2008ag,Alves:2011wf}.}  
constrain the sneutrino-LSP scenario. 
Moreover, we discuss the most promising topologies for which no SMS results exist 
but would enhance the LHC sensitivity to sneutrino DM.
To this aim, we make use of the \smodels\ package~\cite{Kraml:2013mwa,smodels:v1,smodels:wiki} to compare  
the predictions of the MSSM+RN model against the SMS limits published by ATLAS and CMS. 
The strengths of \smodels\  are that it 1.)~automatically decomposes 
the signal of an arbitrary SUSY spectrum into all its SMS-equivalent topologies, and 
2.)~includes a large database of more than 60 SMS results from ATLAS and CMS SUSY searches. 
This allows us to test the limits from a large variety of searches and 
at the same time draw conclusions about which additional topologies should be considered.

The paper is organised as follows. After briefly defining the MSSM+RN in Section~\ref{sec:model} we describe 
the numerical procedure in Section~\ref{sec:num}. In particular in~\ref{sec:sample} we explain the sampling method 
and the constraints implemented in the model likelihood function, while in~\ref{sec:sms} we describe the application 
of \smodels\ to the MSSM+RN. Our numerical results are presented in Section~\ref{sec:result}, and the conclusions 
in Section~\ref{sec:concl}. 
Two appendices contain some interesting supplementary material. 
Appendix~\ref{sec:efficiency} discusses the validity of applying SMS results from slepton searches 
(dilepton signature) to chargino-pair production followed by decays into leptons and sneutrinos.  
Appendix~\ref{sec:lifetime} gives some details on scenarios with long-lived heavy charged particles, 
in particular gluinos or stops, %currently not treated by \smodels.
which so far cannot be constrained by SMS results.

%%%%%%%%%%%%%%%%%%%%%%%%%%%%%%%%%%%%%%%%%%%%%%%%%%%%%%%%%%%%
\section{The MSSM+RN model}\label{sec:model}
%%%%%%%%%%%%%%%%%%%%%%%%%%%%%%%%%%%%%%%%%%%%%%%%%%%%%%%%%%%%

We use the MSSM+RN model as defined in~\cite{Arkani-Hamed:2000bq,Borzumati:2000mc,Arina:2007tm}.  
(The model used in \cite{Belanger:2010cd,Belanger:2011ny,Dumont:2012ee} differs only slightly in notation.)
The superpotential for Dirac RH neutrino superfield is given by
\begin{equation}\label{lrsuppot}
W = \epsilon_{ij} (\mu \hat H^{u}_{i} \hat H^{d}_{j} - Y_{l}^{IJ} \hat H^{d}_{i} \hat L^{I}_{j} \hat R^{J}
+ Y_{\nu}^{IJ} \hat H^{u}_{i} \hat L^{I}_{j} \hat N^{J} )\,,
\end{equation}
where $Y_{\nu}^{IJ}$ is a matrix in flavor space (which we choose to be real and diagonal), from which the mass of neutrinos are obtained as $m_D^{I} = v_{u}Y_{\nu}^{II}$. 
Note that lepton-number violating terms are absent in this scheme. 
The additional scalar fields contribute with new terms in the soft-breaking potential 
\begin{equation}\label{lrsoftpot}
V_{\rm soft} = (M_{L}^{2})^{IJ} \, \tilde L_{i}^{I \ast} \tilde L_{i}^{J} + 
(M_{N}^{2})^{IJ} \, \tilde N^{I \ast} \tilde N^{J} - 
 [\epsilon_{ij}(\Lambda_{l}^{IJ} H^{d}_{i} \tilde L^{I}_{j} \tilde R^{J} + 
\Lambda_{\nu}^{IJ} H^{u}_{i} \tilde L^{I}_{j} \tilde N^{J})  + \mbox{h.c.}]\,,
\end{equation}
where both matrices $M_{N}^{2}$ and $\Lambda_{\nu}^{IJ}$ are real and diagonal, $M_{N}^{2}={\rm diag}(m^2_{N^k})$ and $\Lambda_{\nu}^{IJ}={\rm diag}(A_{\snu}^k)$,
with $k=e,\mu,\tau$ being the flavor index. In the sneutrino interaction basis, defined by the vector $\Phi^\dag=(\tilde{\nu}_L^\ast,\, \tilde N^\ast)$, the sneutrino mass potential is
\begin{eqnarray}
V_{\rm mass}^k =\frac{1}{2}\, \Phi^{\dag}_{LR}\, \mathcal{M}^2_{LR}\, \Phi_{LR}\,,
\end{eqnarray}
with the squared--mass matrix $\mathcal{M}^2_{LR}$ 
\begin{equation}
\mathcal{M}^2_{LR}  =  \left(  \begin{array}{cc}
m^2_{L^k} + \frac{1}{2} m_{Z}^{2} \cos(2\beta) + m_D^2  & \; \; \; \frac{1}{\sqrt{2}} A_{\snu}^k v\sin\beta - \mu m_D/\tan\beta \\
                 \frac{1}{\sqrt{2}} A_{\snu}^k v\sin\beta - \mu m_D/\tan\beta   & m^2_{N^k} + m_D^2
                 \end{array}\right)  \,.
  \label{eq:masslr}
\end{equation}
Here, $m^2_{L^k}$ are the soft mass terms for the three SU(2) leptonic doublets, $\tan\beta = v_u/v_d$ and $v^2=v_u^2+v_d^2=(246\,  {\rm GeV})^2$, with $v_{u,d}$ the usual Higgs vacuum expectation values (vevs).
The Dirac neutrino mass $m_D$ is small and can be safely neglected. 

The off-diagonal term determines the mixing of the LH and RH fields. 
If $A_{\snu}^k = \eta Y_\nu$, that is if the trilinear term is aligned to the neutrino Yukawa, this term is certainly very small as compared to the diagonal entries and is therefore negligible. However, $A_{\snu}^k$ can in general be a free parameter and may naturally be of the order of the other entries of the matrix~\cite{Borzumati:2000mc,Arkani-Hamed:2000bq}, thus inducing a sizable mixing among the interaction eigenstates. The sneutrino mass eigenstates are then given by 
\begin{eqnarray}\label{lr_eigenstates}
\left(\begin{array}{c} \snu_{k_1} \\ \snu_{k_2} \end{array}\right) = 
\left(\begin{array}{cc} -\sin\theta^k_{\snu} &   \cos\theta^k_{\snu} \\ 
                                        \cos\theta^k_{\snu} & \sin\theta^k_{\snu} \end{array}\right) 
\left(\begin{array}{c} \snu_L^k \\ \tilde{N}^k \end{array}\right) \,.
\end{eqnarray}
The relevant parameters at the electroweak (EW) scale for the sneutrino sector
are the two mass eigenvalues $m_{\snu_{k_1}}$ and $m_{\snu_{k_2}}$ and the mixing angle $\theta_{\snu}^k$, related to the $A_{\snu}^k$ term via 
\begin{equation} 
  \sin 2\theta_{\snu}^k = \sqrt{2} \frac{A_{\snu}^k\, v \sin \beta}{(m^2_{\snu_{k 2}}-m^2_{\snu_{k 1}})}\, .
\end{equation}
The sneutrino coupling to the $Z$ boson, %which couples only to left-handed fields, 
which does not couple to SU(2)$_L$ singlets, is largely reduced by a sizeable mixing. This has a relevant impact on the sneutrino phenomenology, as discussed in, {\it e.g.}, Refs.~\cite{Arkani-Hamed:2000bq,Smith:2001hy,Grossman:1997is,Arina:2007tm,Belanger:2010cd}.

The renormalization group equations (RGEs) are modified by the new singlet superfields $\hat{N}$ as
\begin{eqnarray}
\frac{\der m^2_{N^k}}{\der \ln\mu} & = & \frac{4}{16 \pi^2} \left(A^{k}_{\snu}\right)^2\,,\\ \nonumber
\frac{\der m^2_{L^k}}{\der \ln\mu} & = & \left( \rm MSSM\,  terms \right) + \frac{2}{16 \pi^2} \left(A^{k}_{\snu}\right)^2\,,\\  \nonumber
\frac{\der A_{\snu}^k}{\der \ln\mu} & = & \frac{2}{16 \pi^2} \left(- \frac{3}{2} g^2_2 - \frac{3}{10} g_1^2 + \frac{3}{2} Y_t^2 + \frac{1}{2} Y_\tau^2 \right) A_{\snu}^k\,,\\ \nonumber
\frac{\der m^2_{H_u}}{\der \ln\mu} & = & \left( \rm MSSM\,  terms \right) +\sum_{k=e,\mu,\tau} \frac{2}{16 \pi^2} \left(A^{k}_{\snu}\right)^2\,, \nonumber
\end{eqnarray}
with $\mu$ being the renormalization scale, $g_2$ and $g_1$ the SU(2) and U(1) gauge couplings, $Y_{t,\tau}$ the top and $\tau$ Yukawa respectively. Notice that the RH soft mass receives corrections only from the trilinear term, which affects as well the running of the LH part, as recognized in~\cite{Belanger:2010cd,Belanger:2011ny}. 

By neglecting all lepton Yukawas but $Y_\tau$  in the RGEs and by assuming common scalar masses and trilinear couplings for all flavors, the sneutrino tau, $\snu_{\tau_1}$, ends up to be the lightest one among the three sneutrino flavors and hence the LSP, while $\snu_{e_1}=\snu_{\mu_1}$. 
Note that it frequently happens that the mass splitting between $\snu_{\tau_1}$ and $\snu_{e_1,\mu_1}$ is smaller than 5 GeV, which means that regarding collider phenomenology they are practically degenerate. This will be discussed in more detail in Section~\ref{sec:sms}.

%%%%%%%%%%%%%%%%%%%%%%%%%%%%%%%%%%%%%%%%%%%%%%%%%%%%%%%%%%%%
\section{Numerical procedure}\label{sec:num}
%%%%%%%%%%%%%%%%%%%%%%%%%%%%%%%%%%%%%%%%%%%%%%%%%%%%%%%%%%%%

%----------------------------------------------------------------------------------------
\subsection{Sampling method over the model parameters}\label{sec:sample}
%----------------------------------------------------------------------------------------

For definiteness, we study the MSSM+RN with soft terms defined at a high scale
$M \sim M_{\rm GUT}$ as in \cite{Arina:2013zca}. Allowing for
non-universalities in the gaugino and scalar sectors, our set of free
parameters is
\begin{equation}
   M_1, M_2, M_3, m_L, m_R, m_N, m_Q, m_H, A_l,A_{\snu}, A_q, \tan\beta, {\rm sgn}\mu  \,.
\label{eq:param}
\end{equation}
Here the $M_i$ are the gaugino masses, $m_L, m_R, m_N$ are the charged slepton
and sneutrino masses (equal for all flavors), $m_Q$ is a common squark mass
parameter, $m_H\equiv m_{H_u} = m_{H_d}$ denotes the common entry for the two
Higgs doublet masses, and $A_l$ and $A_q$ are the scalar trilinear couplings
for the sleptons and squarks respectively, same for all flavors. The absolute
value of $\mu$ is obtained from the minimization of the Higgs potential, leaving
only the sign of $\mu$ as a free parameter. The computation of the mass spectrum
follows that explained in \cite{Arina:2013zca}, where all details are
provided.
\begin{table}[t!]
\caption{Summary of the observables and constraints used in this analysis.  \label{tab:co}}
\begin{center} \renewcommand\arraystretch{1.2}
\begin{tabular}{| c | c | c | c| }
  \hline
 &  Observable & Value/Constraint  & Ref.\\
  \hline
{\it \underline{Measurements}} &  $m_h$  & $ 125.85 \pm 0.4$ GeV (exp) $ \pm\, 4 $\,  GeV (theo) & ~\cite{atlas:2012gk,cms:2012gu}\\
 &  BR$(B \to X_s \gamma) \times 10^{4}$  & $ 3.55 \pm 0.24 \pm 0.09 $  (exp)  & ~\cite{Aaij:2012nna}  \\
  &  BR$(B_s \to \mu^+ \mu^-) \times 10^{9}$  & $ 3.2^{+1.4}_{-1.2} $  (stat) $^{+0.5}_{-0.3}$\,   (sys) & ~\cite{Amhis:2012bh}  \\ 
 &  $\Omega_{\rm DM} h^2$  & $ 0.1186 \pm 0.0031 $  (exp) $ \pm\, 20\%$\,   (theo) & ~\cite{Ade:2013zuv}  \\
\hline
{\it \underline{Limits}} &   $\Delta\Gamma_Z^{\rm invisible}$  & $ <2 $ MeV  ( 95\% CL) & ~\cite{PDG}  \\ 
&   ${\rm BR}(h \to {\rm invisible})$  & $ <20\% $  (95\% CL) & ~\cite{Belanger:2013xza}  \\ 
 & $m_{\stau}$ & $ > 85$ GeV (95\% CL) & ~\cite{sleptons2004} \\ 
 & $m_{\widetilde{\chi}_1^+}, m_{\tilde{e},\tilde{\mu}}$  &  $>  101$ GeV (95\% CL) & ~\cite{PDG}\\
 & $m_{\tilde{g}}$ & $ > 308$ GeV (95\% CL) & ~\cite{Abazov:2007aa}\\ 
 &   $ \sigma_{n}^{\rm SI}  $  & $ < \sigma^{\rm SI}_{\rm LUX}$ (90\% CL)  & ~\cite{Akerib:2013tjd}\\
  \hline
\end{tabular}
\end{center}\renewcommand\arraystretch{1.0}
\end{table}

The list of constraints implemented in the model likelihood function is given in Table~\ref{tab:co}. 
In particular, besides consistency with  $B$-physics constraints, 
we require the  Higgs mass $m_h$ to be compatible with the ATLAS and CMS
measurements~\cite{atlas:2012gk,cms:2012gu}, which we 
combine by a statistical mean, as obtained in~\cite{Cabrera:2012vu}.  Its uncertainty is dominated by the theoretical error, estimated to be around 4\,GeV~\cite{Allanach:2004rh}. We also require that chargino and charged slepton masses
 fulfill the LEP bounds at 95\% confidence level (CL) ---notice that the tau slepton has a slightly less stringent lower bound of 85 GeV~\cite{sleptons2004} as compared to selectrons and smuons---
and we include the gluino mass bound from the D0 collaboration~\cite{Abazov:2007aa}. 
If $\snu_{\tau_1}$ is light enough to be produced in $Z$ decay, we require its contribution to the $Z$ invisible
decay width to be smaller than 2\,MeV~\cite{ALEPH:2005ab}. Similarly, when the sneutrino mass is lighter than $m_h/2$, the Higgs can decay invisibly into sneutrino pairs. We require that such decays do not contribute more than 20\% to the Higgs invisible branching ratio~\cite{Belanger:2013xza}. 

Regarding DM constraints, we require consistency with the measured relic abundance and with the bounds from direct detection experiments (constraints from indirect DM detection are also fulfilled). 
The experimental error on $\Omega_{\rm DM} h^2$ has become incredibly small due to the Planck measurement~\cite{Ade:2013zuv}, while the theoretical one is still large. We use a conservative  estimate of the order 20\%~\cite{Boudjema:2011ig} for the latter. Furthermore, we enforce the sneutrino SI scattering 
cross section off nuclei, $\sigma_n^{\rm SI}$, to be compatible with the recent 90\% CL bound from LUX~\cite{Akerib:2013tjd}. 

To evaluate the experimental observables we first computed the supersymmetric
particle spectrum with a modified version of
{\sc{SoftSusy}}~\cite{Allanach:2001kg}. For the computation of the sneutrino
relic density and elastic scattering cross-section the model has been
implemented in {\sc{FeynRules}}~\cite{Duhr:2011se,Degrande:2011ua}, by
adding the appropriate term in the superpotential and in the soft SUSY
breaking potential. We generate output files compatible with {\sc{CalcHep}}
in order to use the public code
{\sc{micrOMEGAS\_3.2}}~\cite{Belanger:2013oya}. The $B$-physics
observables are computed by interfacing the program with
{\sc{SuperIso}}~\cite{Mahmoudi:2008tp}. 

The likelihood is constructed in a simple way.  For measured quantities, we
assume a Gaussian likelihood function with a variance given by combining in
quadrature the theoretical and experimental variances. For observables for
which only lower or upper limits are available, we use a likelihood modelled
as a step function on the $x\%$ CL of the exclusion limit.
The total likelihood function is then the product of the individual
likelihoods associated to each experimental result.  In order to save time in
the sampling procedure, the slepton, chargino and gluino mass limits are,
however, absorbed into the prior probability density functions: each parameter
point generating a mass spectrum that violates one of these bounds is
immediately discarded.

Given the likelihood function, we sample the posterior probability density
function with the {\sc{MultiNest}}
algorithm~\cite{Feroz:2007kg,Feroz:2008xx,2013arXiv1306.2144F}.  In order to
cover all phenomenological interesting classes, we run separate chains that
look either for light(ish) EW-inos ($m_{\tilde{\chi}_1^{\pm}} < 900$ GeV),
light sleptons ($m_{\tilde{l}} < 600$ GeV), or for light squark or gluinos
($m_{\tilde{q}} < 1.5$ TeV or $m_{\tilde{g}} < 1.5$ TeV).  As for the choice
of priors, we always take logarithmic priors on $M_3, m_Q, A_Q, m_H$, while we
use both logarithmic and flat priors for $M_1, M_2, m_L, m_R,m_N,A_L,A_{\snu},
\tan\beta$, the sign of $\mu$ is fixed to +1 (details on the prior ranges are
provided in~\cite{Arina:2013zca}). 
In particular we perform two chains, one with log and one with flat priors,
for each relevant data set: two chains for light EW-inos (these two data sets
coincide with the ones used in~\cite{Arina:2013zca}), two chains for light
sleptons and two chains for light squarks or gluinos. In each case, the other
masses are left to vary freely from high to low values. The motivation for
this is, as mentioned, to cover all potentially interesting cases; the results
we will present in Section~\ref{sec:result} are for all chains combined
together.

The sampled points correspond to a 95\%~CL in volume of the posterior. (Since in this study we are not interested 
in statistical statements on the parameter space, we will however not exploit this feature.)
The limits imposed by a step function 
are of course strictly obeyed by all scan points. Moreover, we have checked that none of the individual constraints implemented by a Gaussian gets a large pull in the final sample. In particular, BR$(B \to X_s \gamma)$ and BR$(B_s \to \mu^+ \mu^-)$  are in full agreement with the 95\% CL experimental results~\cite{Aaij:2012nna,Amhis:2012bh} for all points in the samples. 

Once the sampling of the parameter space according to the constraints in Table~\ref{tab:co} is completed, all the points in the chains are confronted against the LHC Run 1 results by means of \smodels\ as explained in the next subsection.

\clearpage
%----------------------------------------------------------------------------------------
\subsection{Deriving LHC constraints with SModelS}\label{sec:sms}
%----------------------------------------------------------------------------------------

\smodels~\cite{Kraml:2013mwa,smodels:v1,smodels:wiki} is designed to decompose the signal of any 
arbitrary BSM spectrum with a $\mathbb{Z}_2$ symmetry 
into simplified model topologies and test it against the existing LHC bounds in the SMS context. 
\smodels\ uses Pythia\,6.4~\cite{Sjostrand:2006za}, NLL-fast~\cite{Beenakker:1996ch,Beenakker:1997ut,Kulesza:2008jb,Kulesza:2009kq,Beenakker:2009ha,Beenakker:2010nq,Beenakker:2011fu} and PySLHA~\cite{Buckley:2013jua}, and includes a database of more than 60 SMS results from ATLAS and CMS. 
The decomposition procedure works ``out of the box'' for the MSSM+RN model with a sneutrino LSP. 
Nonetheless some subtleties must be taken care of when processing the MSSM+RN scan points with \smodels.  

First, the input to \smodels\ can be simulated events or an SLHA~\cite{Skands:2003cj} file containing 
the full mass spectrum and decay tables as well as the SUSY production cross sections, $\sigma$,  
in the format  specified at~\cite{slha-xsext}. We choose the latter option. 
We use the MSSM+RN model implemented in {\sc micrOMEGAs}\,3.2~\cite{Belanger:2013oya} (see \cite{Arina:2013zca}) 
to compute the decay branching ratios, $\br$.  The production cross sections for sleptons and sneutrinos 
(\ie\ the sector modified with respect to the MSSM) are also computed with {\sc micrOMEGAs}\,3.2. 
For all other production processes, we use the default \smodels\ 
cross section calculator based on  Pythia\,6.4~\cite{Sjostrand:2006za} and NLL-fast~\cite{Beenakker:1996ch,Beenakker:1997ut,Kulesza:2008jb,Kulesza:2009kq,Beenakker:2009ha,Beenakker:2010nq,Beenakker:2011fu}. 
Electroweak cross sections are thus computed at leading order while strong productions are computed at NLO+NLL order. 

Given the information on $\sigma$ and $\br$ in the SLHA files, \smodels\ computes $\sigmaXBF$ for each topology that occurs. Here, a topology is characterised by the SM particles originating from each vertex, and the mass vector of the SUSY  particles in the decays. In order to avoid dealing with a large number of irrelevant processes, which is expensive in terms of computing time, topologies for which $\sigmaXBF < \sigma_{\rm cut}$, with $ \sigma_{\rm cut}=0.05$~fb, are discarded.  

\begin{figure}[t]\centering
\includegraphics[width=0.5\textwidth]{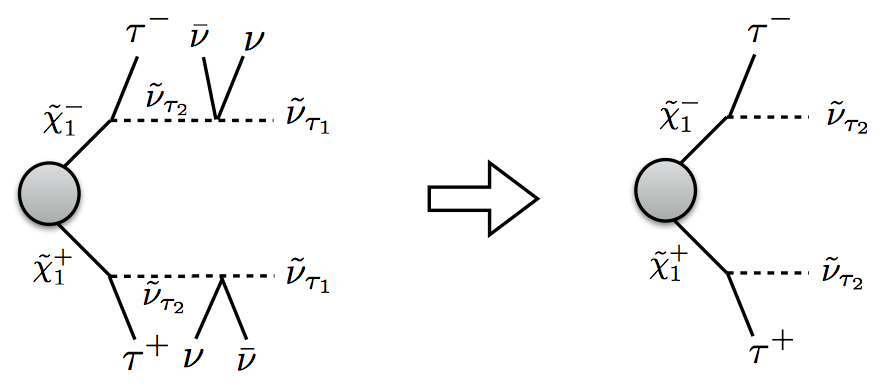} 
\caption{Illustration of ``invisible compression'' in \smodels. The decays of the heavier sneutrino to the lighter one plus neutrinos are discarded in the final topology, leaving the $\tilde\nu_{\tau_2}$ as an effective LSP.}
\label{fig:invisible-illustration} 
\end{figure}

When dealing with an arbitrary spectrum of SUSY particles, it is possible that a part of the decay chain leads to completely invisible decays,  \eg\ a decay of a heavy sneutrino to a lighter one plus neutrinos in the current scenario. 
In such cases, \smodels\ compresses the invisible part of the decay chain as illustrated in Fig.~\ref{fig:invisible-illustration}. 
All decays to neutrinos appearing after the last visible decay are disregarded, yielding an ``effective LSP'' for the particular event, which can be different from the true LSP. This procedure is called ``invisible compression''. Likewise, a neutralino may decay invisibly to a sneutrino and a neutrino; in this case the compressed topology resembles an MSSM topology.

In addition, if the mass gap between mother and daughter particles is small, the decay products will be too soft to be detected at the LHC. This is taken care of by the so-called ``mass compression'' in \smodels, discarding any SM particle that come from a vertex for which the mass splitting of the R-odd particles is less than a certain threshold. We use $5$ GeV as the minimum required mass difference for the decay products to be visible. 

Another comment is in order.  
The experimental constraints currently implemented in the \smodels\ database require final states containing 
missing transverse energy (MET). 
This means that scenarios with long-lived particles ($c\tau>10$~mm) leading to signatures with displaced vertices 
or heavy charged particle tracks cannot be tested with \smodels. 
In the MSSM, this problem occurs, \eg, in wino-LSP scenarios where the ${\widetilde{\chi}_1^\pm}$ is 
highly mass-degenerate with the ${\widetilde{\chi}_{1}^0}$ and thus becomes long-lived. 
In the sneutrino LSP case, not only charginos can be long lived if the mass splitting with the sneutrino is  small enough; 
other possibilities are, \eg, long-lived gluinos or stops, if they are the next-to-lightest SUSY particle (NLSP). 
We perform a detailed check of all input points to avoid the erroneous application of SMS limits to such cases. 
Points that have visible decays from long-lived particles or heavy charged particle tracks with cross sections 
larger than $\sigma_{\rm cut}$ are discarded.  
(A brief discussion of such scenarios can be found in Appendix~\ref{sec:lifetime}.)

Once the decomposition into SMS topologies, including mass and invisible compression, is completed and 
the checks that the SMS results actually apply are passed, a given point is confronted against the SMS results in the \smodels\ database. For each experimental constraint that exists, \smodels\ reports among other things the 
analysis name, the {\bf Tx} name identifying the topology,\footnote{The {\bf Tx} names are explained in the SMS dictionary on %\href{http://smodels.hephy.at/wiki/SmsDictionary}{http://smodels.hephy.at/wiki/SmsDictionary}.
http://smodels.hephy.at/wiki/SmsDictionary} 
the predicted signal cross section for the point under consideration and the 95\% CL experimental upper limit on it. 
Finally, the ratio $r$ of the signal cross-section and the upper limit, 
$r=\sigma(\textrm{predicted})/\sigma(\textrm{excluded})$, is given, where $\sigma$ effectively means 
$\sigmaXBF$ or the weight of the topology. 
A value of $r\ge 1$ means that the input model is likely excluded by the corresponding analysis.

%%%%%%%%%%%%%%%%%%%%%%%%%%%%%%%%%%%%%%%%%%%%%%%%%%%%%%%%%%%%
\section{Results}\label{sec:result}
%%%%%%%%%%%%%%%%%%%%%%%%%%%%%%%%%%%%%%%%%%%%%%%%%%%%%%%%%%%%

We now turn to analysing the impact of the LHC searches on the MSSM+RN parameter space. 
As explained in the previous Section, we here consider only points for which the SMS results apply, 
\ie\ we discard points with non-prompt visible decays as well as points with long-lived charged particles. 
Scanning over the parameter space, we can then distinguish several cases: 
\begin{itemize}
\item the SMS results in principle apply but no SMS constraints actually exist for the specific topologies of the point --- these points will be labelled as {\em not tested};\footnote{This occurs if no simplified model result exists for the signal topologies of the point considered, but also if the mass vector of a topology lies outside that of the experimental constraint. Moreover, we include here also the points for which all signal topologies are discarded because of $\sigmaXBF < \sigma_{\rm cut}$.}
\item there exist (one or more) SMS results that test the specific topologies of the point but for each topology the total $\sigma \times \br$ is below the corresponding 95\% CL upper limit --- these points will be considered as {\em allowed}; and
\item at least one topology has a  $\sigma \times \br$  equal or above its 95\% CL upper limit ($r\ge1$) --- these points will be considered as {\em excluded}.
\end{itemize}

\begin{figure}[t!]\centering
\includegraphics[height=7cm]{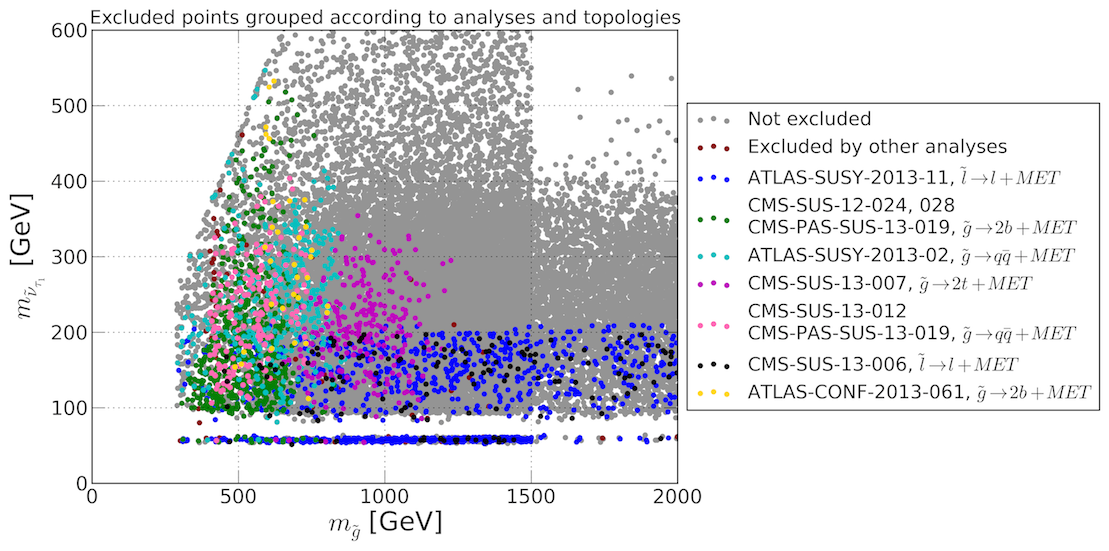}
\caption{For scan points that are excluded by the SMS limits, we show (in color) the breakdown of most constraining analyses in the $\snu_{\tau_1}$ vs.\ $\tilde g$ mass plane. To illustrate the coverage of the parameter space, we also show (in grey) the not excluded or not tested points.}
\label{fig:breakdown-snu-gluino} 
\end{figure}

\begin{figure}[t!]\centering
\includegraphics[height=7cm]{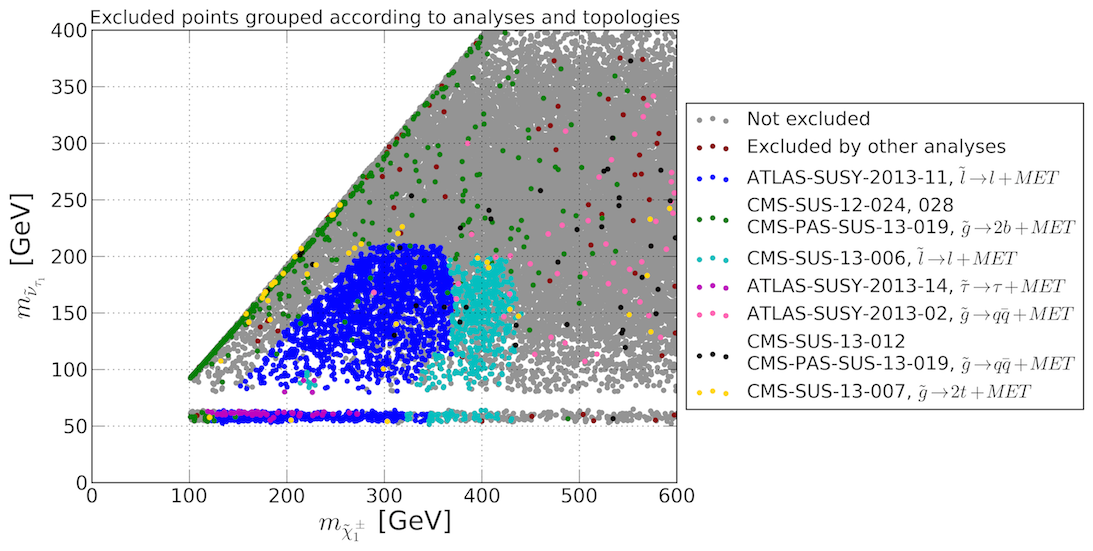}\caption{As Fig.~\ref{fig:breakdown-snu-gluino} but  in the $\snu_{\tau_1}$ vs.\ $\cha_1$ mass plane.}
\label{fig:breakdown-snu-c1} 
\end{figure}

Let us start with the question which analyses are the most important ones for constraining the model. 
To this end, Figs.~\ref{fig:breakdown-snu-gluino} and \ref{fig:breakdown-snu-c1} show a breakdown 
of most constraining analyses in the 
$\snu_{\tau_1}$ versus $\tilde g$ and $\snu_{\tau_1}$ versus $\cha_1$ mass planes, respectively. 
Looking first at Fig.~\ref{fig:breakdown-snu-gluino}, we see that (the SMS interpretations of) 
the hadronic SUSY searches \cite{Aad:2013wta,Aad:2014wea,ATLAS-CONF-2013-061,Chatrchyan:2013wxa,Chatrchyan:2013lya,Chatrchyan:2013iqa,Chatrchyan:2014lfa} 
are constraining gluino masses up to about $m_{\tilde g}\approx 1200$~GeV and LSP masses up to 
about $m_{\snu_{\tau_1}}\approx 500$~GeV. These searches mostly exclude points where either 
$\tilde g\to b\bar b\neu_i$, $\tilde g\to t\bar t\neu_i$ or $\tilde g\to q\bar q\neu_i$ decays are dominant, 
followed by an invisible decay of the neutralino, $\neu_i\to \nu\snu$. 
Moreover,  dilepton + MET searches~\cite{Aad:2014vma,Khachatryan:2014qwa} 
exclude sneutrino LSP masses up to  about $m_{\snu_{\tau_1}}\approx 210$~GeV, independent of the gluino mass. The process that is constrained here is Drell-Yang production of $\chap_1\cham_1$ 
followed by $\cha_1\to l^\pm \snu_{l1}$ ($l=e$ or $\mu$), with the  $\snu_{l1}\to\snu_{\tau_1}+X$ decay being invisible (because of $X$ being genuinely invisible or very soft). Consequently, in Fig.~\ref{fig:breakdown-snu-c1} we see that chargino masses can be excluded up to about  $m_{\cha_1}\approx 440$~GeV by the dilepton + MET limits. (There is 
also a small region of parameter space at low masses where $\tau^+\tau^- + {\rm MET}$~\cite {Aad:2014yka} 
gives the strongest limit.)

It is important to note here that the constraints on $\chap_1\cham_1\to l^+l^- + {\rm MET}$ actually stem from the
$\tilde l^+\tilde l^- \to l^+l^-\neu_1\neu_1$ simplified model (and analogously for $\tau^+\tau^- + {\rm MET}$), 
which has the opposite spin configuration than chargino-pair production followed by chargino decays into sneutrinos.  
The validity of applying the limits from the slepton searches to the case of chargino-pair production is discussed in Appendix~\ref{sec:efficiency}.

\begin{figure}[t!]\centering
\includegraphics[width=0.44\textwidth]{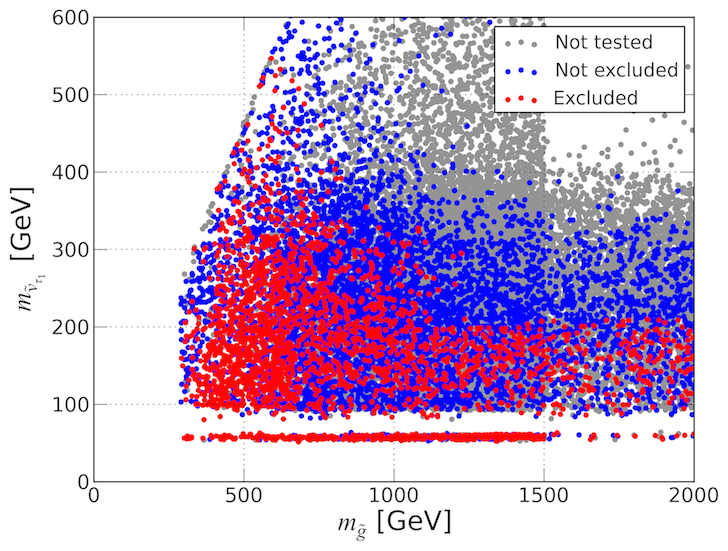}
\includegraphics[width=0.44\textwidth]{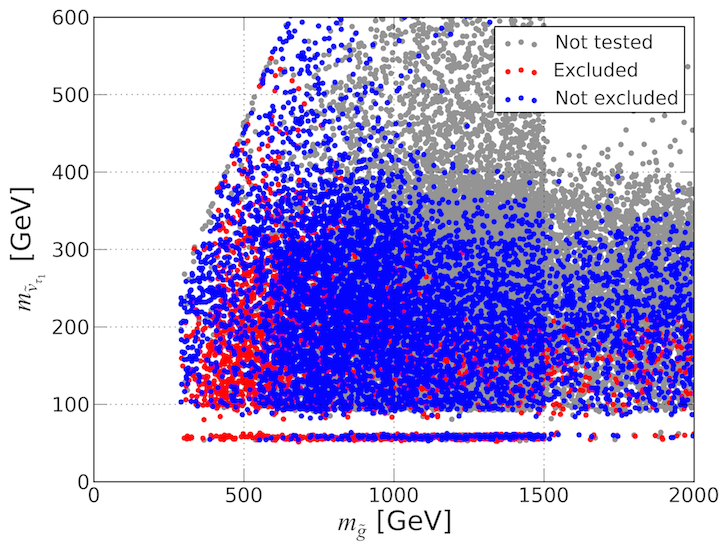} 
\includegraphics[width=0.44\textwidth]{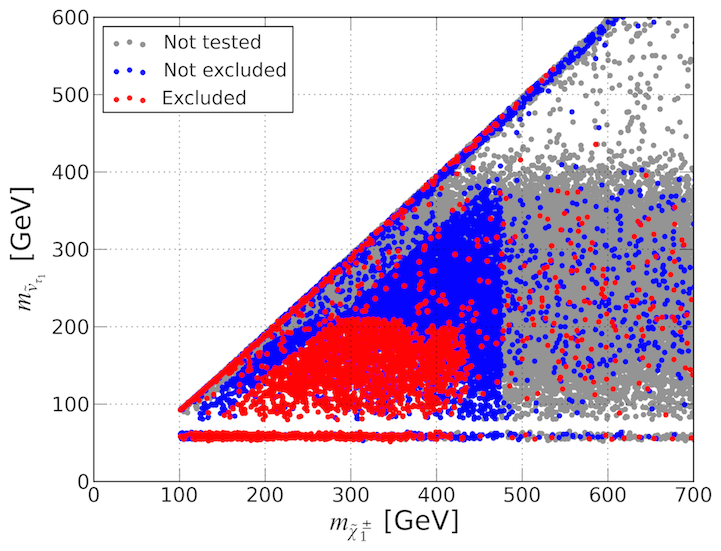}
\includegraphics[width=0.44\textwidth]{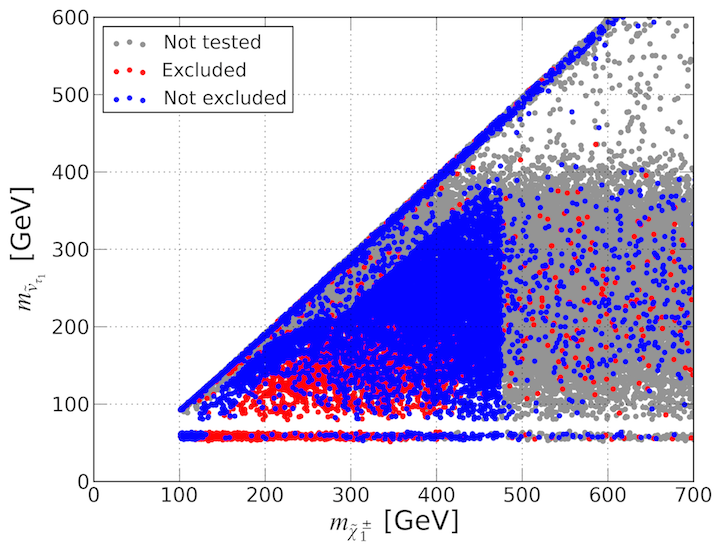} 
\caption{Scatter plots of points for which SMS results apply. 
The top row shows the $\snu_{\tau_1}$ vs.\ $\tilde g$, the bottom row the $\snu_{\tau_1}$ vs.\ $\cha_1$ mass plane. 
In the panels on the left, the points excluded by the SMS constraints (red) are plotted on top of those which are not excluded (blue); in panels on the right this plotting order is inverted. 
Also shown (in grey) are the ``not tested'' points, for which no SMS constraints exist.}
\label{fig:summary} 
\end{figure}

Also noteworthy is the fact that most of the excluded points in Figs.~\ref{fig:breakdown-snu-gluino} and \ref{fig:breakdown-snu-c1} have some grey points lying below them, which are not excluded or not tested at all. This is corroborated in 
Fig.~\ref{fig:summary}, where we present the summary of not tested, allowed and excluded points in the $\snu_{\tau_1}$ versus $\tilde g$  and $\snu_{\tau_1}$ versus $\cha_1$ mass planes.  In the plots on the left, the excluded points (red) are plotted on on top of the allowed points (blue), while in the plots on the right this plotting order is inverted. 
As can be seen, only a small part of the parameter space can genuinely be excluded by the SMS results---over most 
of the  regions where the SMS results are valid, there are almost always parameter combinations such that the limits 
can be avoided. 

\begin{figure}[t!]\centering
\includegraphics[height=7cm]{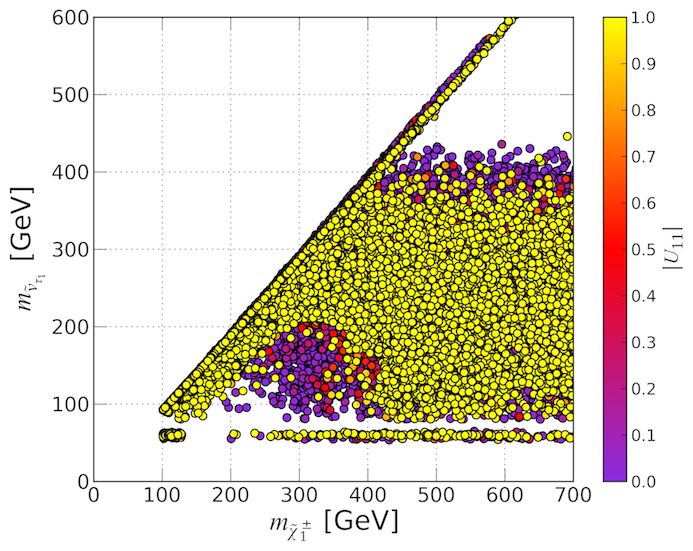}
\caption{Allowed points in the $\snu_{\tau_1}$ vs.\ $\cha_1$ mass plane, with the color code indicating the wino/higgsino content of the $\cha_1$ ($|U_{11}|=1$ means a pure  wino while $|U_{11}|=0$ means a pure higgsino). }
\label{fig:charginomixing} 
\end{figure}

For the dilepton signature originating from chargino-pair production, the chargino mixing plays an important r\^ole: wino-like charginos have a higher production cross section, and a higher branching fraction into $l\snu_{l1}$. The limits from $l^+l^-+{\rm MET}$ searches therefore mostly affect scenarios with wino-like $\cha_1$, while higgsino scenarios are much less constrained. For illustration see Fig.~\ref{fig:charginomixing}, which shows the SMS-allowed points in the $\snu_{\tau_1}$ versus $\cha_1$ mass plane---here the color map gives the size of the $U_{11}$ entry of the chargino mixing matrix,  indicating to the wino/higgsino content of the $\cha_1$. As can be seen, in the region that is in principle constrained by the SMS results the surviving points feature $\cha_1$s that have a large higgsino admixture ($|U_{11}|\lsim0.5$). These points have a lower $\chap_1\cham_1$ production cross section and the $\cha_1$ decays preferably into $\tau\snu_{\tau_1}$ since the higgsino decay to $e,\,\mu$ is Yukawa suppressed;  $\tau^+\tau^-+{\rm MET}$ is however a more difficult signature experimentally and thus only constrains a small strip at low $\snu$ mass, cf.\ the purple points in Fig.~\ref{fig:breakdown-snu-c1}.

%--------------------------------------------------
\subsection*{Missing topologies}
%--------------------------------------------------

The next question to ask is which are the most important signatures not covered by SMS results. 
Such information can be used to improve on the interpretation of the LHC searches for new physics. 
We call these uncovered signatures ``missing topologies''. For any point passed through \smodels, we keep up to ten missing topologies sorted by their $\sigma \times \br$. 
To avoid double counting, missing topologies are evaluated after mass and invisible compressions. 
The total weight is computed by summing over all diagrams giving the same topology,  \ie\ ignoring the mass 
vector of the SUSY states involved.
Moreover, $l=e, \mu$ lepton flavors appearing in the final state are summed over 
(light quark flavors are always summed over). 
In the following, we only consider MSSM+RN scan points which are not excluded, and we demand 
that missing topologies have $\sigma \times \br\ge 1$~fb. 
The results can be presented in two ways, either by showing the most frequent missing topologies 
in a certain parameter space, or by selecting for each parameter point the missing topology 
with the highest cross section. 

\begin{figure}[t!]\centering
\includegraphics[height=7cm]{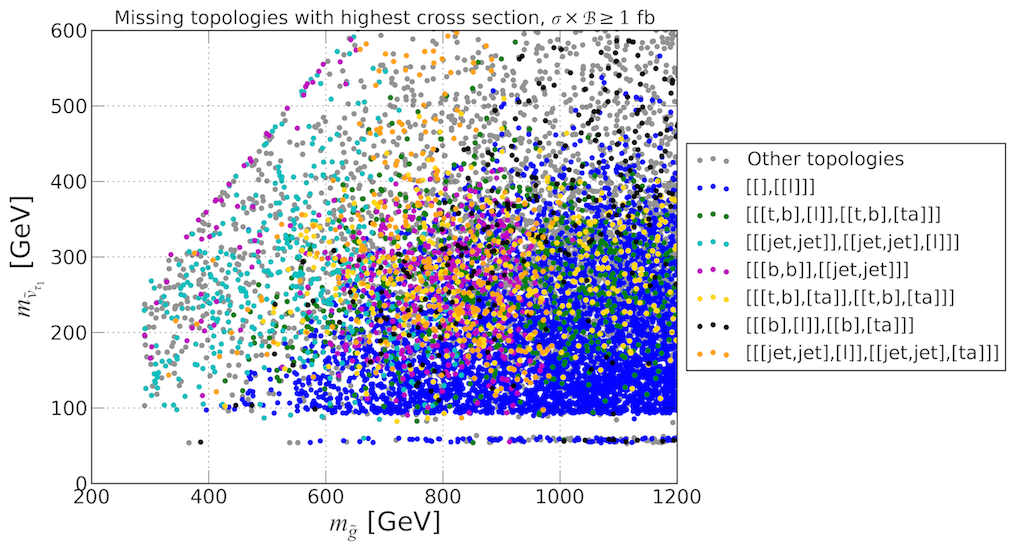}
\caption{Missing topologies with highest $\sigma \times \br$ in the $\snu_{\tau_1}$ vs.\ $\tilde g$ mass plane.}
\label{fig:missingtopos-gluino} 
\end{figure}

We choose the latter approach to show in Fig.~\ref{fig:missingtopos-gluino} the missing topologies in the 
sneutrino- vs.\ gluino-mass plane. 
The various processes are denoted in the bracket notation of \smodelsnn, explained in \cite{Kraml:2013mwa}. 
The structure is {\tt [branch1, branch2]} for the decay chains (``branches'') of the two initially produced 
SUSY particles; each branch contains inner brackets for each vertex, containing in turn the lists of outgoing 
standard model particles. 
Thus, {\tt [[[b,b]],[[jet,jet]]]} denotes gluino-pair production with one gluino decaying into $b\bar b$ + MET 
(via $\tilde g\to b\bar b\neu$, $\neu\to\nu\snu$) and the other one into $q\bar q$ + MET.  
Likewise, {\tt [[[jet,jet]],[[jet,jet],[l]]]} denotes gluino-pair production with the first gluino decaying via 
$\tilde g\to q\bar q\neu$, $\neu\to\nu\snu$ and the other one via $\tilde g\to q\bar q'\cha$, $\cha\to l\snu$.\footnote{More generically, {\tt [[[b,b]],[[jet,jet],[l]]]} denotes production of $XY$ with $X$ undergoing a 1-step decay chain, $X\to b\bar b$ + MET ({\tt branch1=[vertex1]=[[b,b]]}) and $Y$ undergoing a 2-step decay chain, $Y\to q\bar q+Z \to q\bar q+l+{\rm MET}$  ({\tt branch2=[vertex1,vertex2]=[[jet,jet],[l]]}); $X$ can be different from $Y$ or both can be the same.} 

It is apparent that many points with gluino masses below about $1.2$~TeV, for which the LHC searches should have good sensitivity, are not excluded by the SMS results because they feature ``mixed topologies'', where the two pair-produced gluinos undergo different decays (\eg\ one gluino decaying into $b\bar b$ and the other one into light jets). 
Since the SMS results for pair-produced sparticles always assume two identical branches, these cases cannot be constrained by \smodels. Moreover, hadronic final states with additional leptons, as they arise from gluino decays into charginos and the chargino decaying further into a charged lepton ($e,\mu$ or $\tau$) plus the LSP, 
do not have any SMS equivalent. 
Finally, there are no SMS results available for $\tilde g \to tb \cha_j$, no matter of whether the chargino has any visible decays. 

\begin{figure}[t!]\centering
\includegraphics[height=7cm]{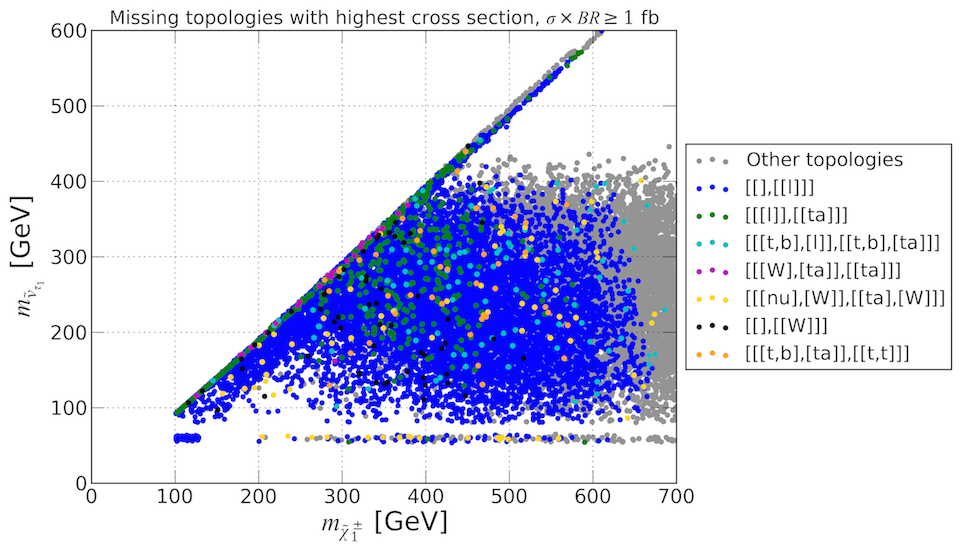}
\caption{Missing topologies with highest $\sigma \times \br$ in the $\snu_{\tau_1}$ vs.\ $\cha_1$ mass plane.}
\label{fig:missingtopos-chargino} 
\end{figure}

\begin{figure}[t!]\centering
\includegraphics[height=7cm]{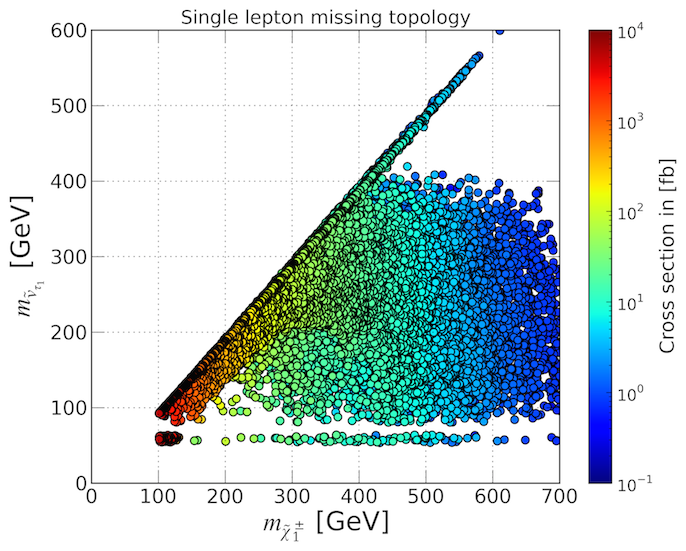}
\caption{Cross sections $\sigma \times \br$ for the single lepton + MET missing topology for not excluded or not tested points.}
\label{fig:singlelepton} 
\end{figure}

It is also worth noting that over a large part of the parameter space single lepton + MET ({\tt [[],[[l]]]} in bracket notation) 
is the most important missing topology. This signature arises from $\neu_i\cha_j$ production; its importance is 
corroborated in Fig.~\ref{fig:missingtopos-chargino}, where one can see that it is indeed dominating the whole 
sneutrino- vs.\ chargino-mass plane. (There are also cases where single $W$ + MET is dominant.)
The cross section for single lepton + MET production, shown in Fig.~\ref{fig:singlelepton}, can be very large and 
should give important additional constraints on the model. 
While searches for single lepton + MET were performed by both ATLAS~\cite{ATLAS:2014wra} and
CMS~\cite{Khachatryan:2014tva}, unfortunately no suitable SMS interpretation exists for these analyses. It would be extremely interesting if the experimental collaborations provided upper limit maps and/or efficiency maps for their single lepton + MET analyses in the context of a chargino--sneutrino simplified model. 

Having both light EW-inos and light staus can generate decay chains with more `exotic' 
signatures, in particular $\cha_i\neu_j$ followed by 
$\cha_i \to \nu \tilde{\tau}^{\pm} \to \nu W^{\pm}\snu_{\tau_1}$ and 
$\neu_j \to \tau^{\pm} \tilde{\tau}^{\mp} \to \tau^{\pm} W^{\mp}\snu_{\tau_1}$. 
This appears as {\tt [[[nu],[W]],[[ta],[W]]]} (yellow points) in Fig.~\ref{fig:missingtopos-chargino} and is interesting because the $\neu_j$ decay produces with the same rate $\tau^+W^-$ and $\tau^-W^+$: together with the chargino decay this gives rise to a same-sign $W$ signature, $W^\pm W^\pm \tau^\mp$ + MET. 

\begin{figure}[t!]\centering
\includegraphics[height=7cm]{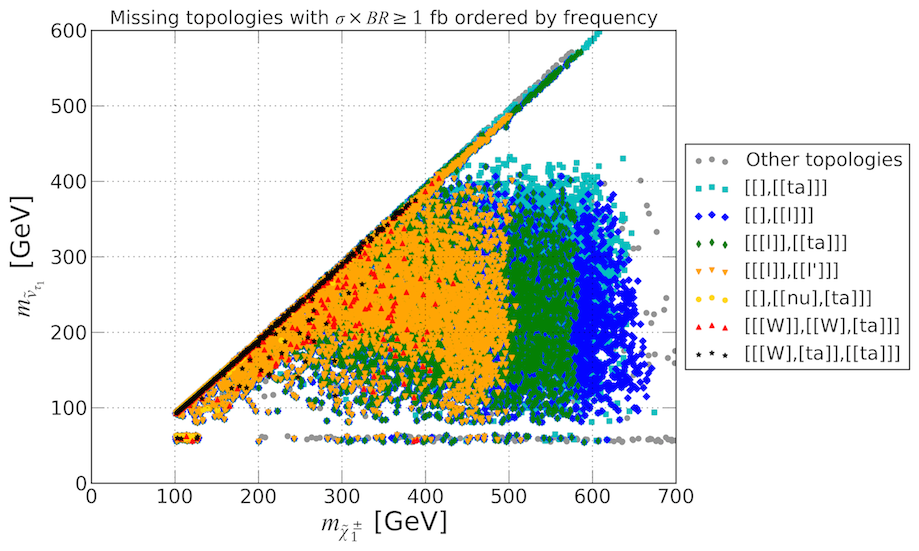}
\caption{Missing topologies with $\sigma \times \br\ge 1$~fb in the sneutrino- vs.\ chargino-mass plane ordered by  frequency of occurrence. The ordering is from top to bottom in the legend, with single tau being the most frequent missing topology, followed by single lepton ($l=e,\mu$), lepton--tau, and so on. ``Other topologies'' are shown on top of the legend without considering their total count (however, each single one of them is less frequent than any of the topologies denoted explicitly).}
\label{fig:missingtopos-EW} 
\end{figure}
 
Before proceeding it is instructive to take another look at the missing topologies arising from EW-ino and slepton production, but this time ordered by their frequency of occurrence. This is done in Fig.~\ref{fig:missingtopos-EW}. 
Not surprisingly we see that besides single lepton ($e$ or $\mu$), single $\tau$ is an important signature. Although it is less clean experimentally, the relative weight of single $e,\mu$ or $\tau$ + MET might potentially give information on the mass pattern of the mostly RH sneutrinos. 
Another important class of ``missing topologies'' are different-flavor dileptons ({\tt [[[l]],[[ta]]]} and {\tt [[[l]],[[l']]]}). 
Different-flavor dileptons + MET have in principle been considered by ATLAS and CMS in the context of 
chargino-pair production in the MSSM with the charginos decaying either into $W^{(*)}\neu_1$~\cite{Aad:2014vma} 
or into $l\nu\neu_1$ via on-shell sleptons/sneutrinos~\cite{Aad:2014vma,Khachatryan:2014qwa}. 
However, the associated SMS limits do not apply to the sneutrino LSP case for various reasons. 
For example, the leptons from $\cha_1\to W^{(*)}\neu_1$ are generally softer than those from 
$\cha_1\to l^\pm\snu_l$ decays (for the same $\cha_1$ and LSP masses) because of the 
additional neutrinos in the $W$ decay. 
The limits for the $\chap_1\cham_1\to 2\times \tilde l\nu ({\rm or}~\tilde\nu l)\to 2\times l\nu\neu_1$ simplified model 
are also not applicable because they involve an additional intermediate mass scale. 

Finally, the {\tt [[[W]],[[W],[ta]]]} topology again gives rise to same-sign $W$'s, see the red triangles in Fig.~\ref{fig:missingtopos-EW}. 
Similarly it is possible to have same sign $\tau$'s arising from {\tt [[[W],[ta]],[[ta]]]} (black stars). In this case, after   $\neu_i\cha_j$ production, the decay chain is $\neu_i \to W^{\mp} \cha_k \to \ W^{\mp} \tau^{\pm}\snu_{\tau_1}$ and $\cha_j \to \tau^{\pm} \snu_{\tau_1}$.

\clearpage
%--------------------------------------------------
\subsection*{Complementarity with direct DM searches}
%--------------------------------------------------

\begin{figure}[t!]\centering
\includegraphics[height=7cm]{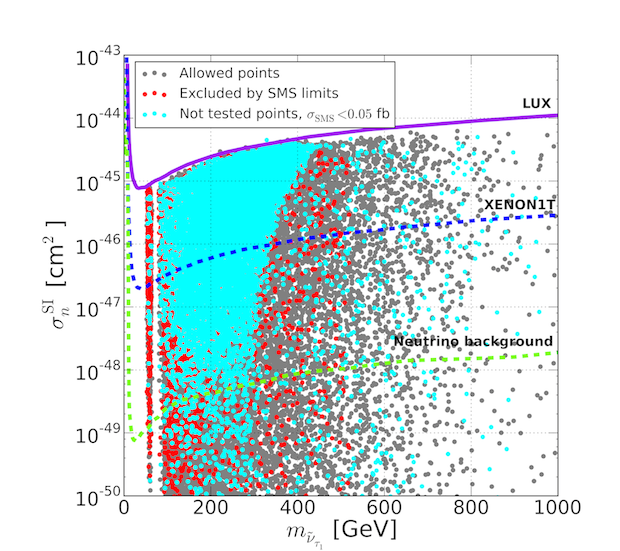}\includegraphics[height=7cm]{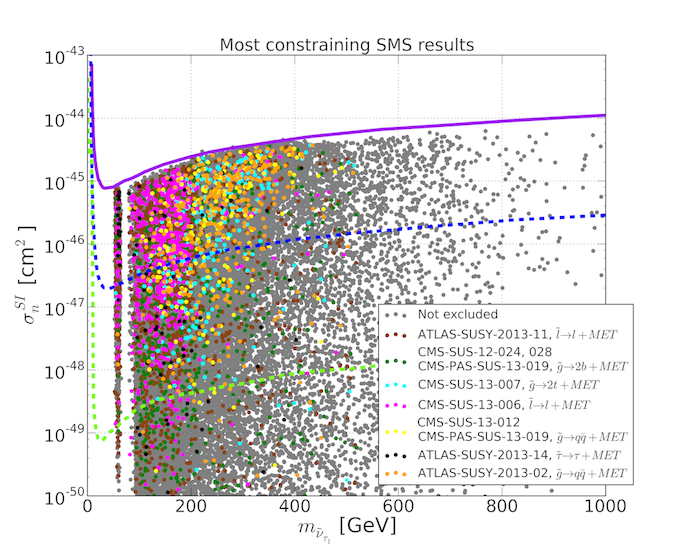}
\caption{Complementarity of LHC and direct DM detection experiments. 
The panel on the left shows SMS allowed, excluded and not tested points in the plane of $\sigma^{\rm SI}_n$ vs.\ $m_{\snu_{\tau_1}}$. The panel on the right shows the breakdown of most constraining analyses 
for the points that are excluded by the SMS limits (for the sake of comparison, the allowed points are shown in grey). In both panels, the solid magenta lines and the dashed blue lines are the current exclusion limit by LUX and the forecasted sensitivity of XENON1T experiment respectively, while the dashed light green line corresponds to the predicted neutrino coherent scattering on nuclei.}
\label{fig:complementarity} 
\end{figure}

\begin{figure}[t!]\centering
\includegraphics[height=7cm]{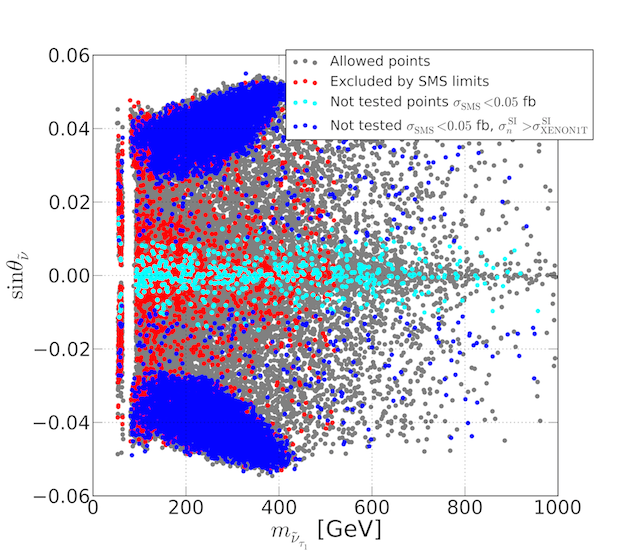}\includegraphics[height=7cm]{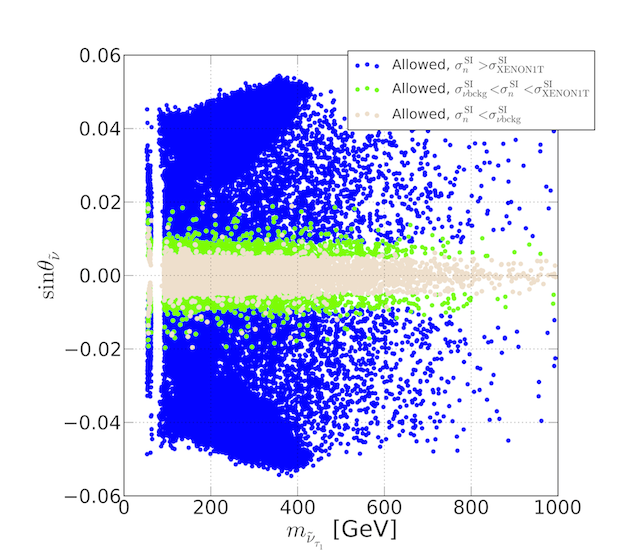}
\caption{On the left allowed (grey), excluded (red) and not tested (blue and cyan) points are shown in the plane of sneutrino mass versus mixing angle. The subset of points with exceedingly small $\sigma\times\br$ at the LHC but in the reach of XENON1T is visualised in blue. On the right, we show the SMS allowed points in the sneutrino mass versus mixing angle 
plane, subdivided in blue points, which are in the reach of XENON1T and in light green (light grey) points, which are above (below) the neutrino background.}
\label{fig:complementarity1} 
\end{figure}

Let us finally turn to the complementarity of LHC and direct DM searches---recall that all points in our scans are consistent with DM constraints, as described in Table~\ref{tab:co}. In Fig.~\ref{fig:complementarity}, left panel, we plot the allowed (gray), excluded (red) and not tested points (cyan) as a function of the sneutrino mass and the SI scattering cross section. In the same plot we also show the forecasted sensitivity of XENON1T after two years of scientific run~\cite{Aprile:2012zx} and the predicted value for neutrino coherent scattering on nuclei~\cite{Billard:2013qya}, which can be an irreducible background for direct detection experiments. From this plot, the complementarity between the two type of searches is striking. Points with a SI elastic cross section well below the neutrino background, and hence not detectable by direct detection experiments, are already excluded by SMS results. On the other hand, a bulk of points allowed (or even more interestingly, not tested) by SMS results is well in the reach of XENON1T, expected to start running in 2015. Notice however that there still exist combinations of parameters that allow sneutrino DM to escape both direct detection and LHC searches, represented by the cyan points below the neutrino background curve. In the MSSM+RN, DM direct searches are basically sensitive to the mass of the LSP and its couplings with the Higgs and $Z$ bosons. The rest of the SUSY mass spectrum is not relevant. This is different with respect to the MSSM with the neutralino LSP, where the interaction with the quarks is mediated as well by squarks on $t$-channel. This is clearly visible in the right panel of Fig.~\ref{fig:complementarity}, which shows the most constraining SMS analyses. In Figs.~\ref{fig:breakdown-snu-gluino} and~\ref{fig:breakdown-snu-c1} these SMS analyses are typically correlated with the gluino or chargino mass, while now they are scattered all over the $\sigma^{\rm SI}_n$  versus $m_{\snu_{\tau_1}}$ plane.

The same set of allowed, excluded and not tested points are plotted as a function of the sneutrino mixing angle in the left panel of Fig.~\ref{fig:complementarity1}.  The bulk of not tested points in the reach of XENON1T (dark blue points) 
has, as expected, relatively large mixing angles, corresponding to sizeable contributions from $Z$ boson exchange to the SI scattering cross section. Excluded red points are scattered everywhere in the $\sin\theta_{\snu}$ vs.\ $m_{\snu_{\tau_1}}$  plane and probe also very RH sneutrinos. In the right panel of Fig.~\ref{fig:complementarity1} we see that among the allowed points, XENON1T can constrain a large portion of the sneutrino parameter space, while the very RH sneutrinos will remain inaccessible to future direct detection detectors. In general the points with negligible mixing angles have $\snu_{\tau_1}$ as LSP and the neutralino as NLSP, which tends to be almost degenerate with chargino. The relic density is then actually achieved by co-annihilation of neutralino-chargino and then communicated to the mostly sterile LSP (see~\cite{Arina:2013zca} for details). Such scenarios are very difficult to test.

\clearpage
%%%%%%%%%%%%%%%%%%%%%%%%%%%%%%%%%%%%%%%%%%%%%%%%%%%%%%%%%%%%

%%%%%%%%%%%%%%%%%%%%%%%%%%%%%%%%%%%%%%%%%%%%%%%%%%%%%%%%%%%%
\section{Conclusions}\label{sec:concl}
%%%%%%%%%%%%%%%%%%%%%%%%%%%%%%%%%%%%%%%%%%%%%%%%%%%%%%%%%%%%

Scenarios with a sneutrino as the LSP are an interesting alternative to MSSM models with neutralino LSPs. Indeed in SUSY models with a RH neutrino superfield (MSSM+RN) the fermionic field contributes to neutrino masses while the scalar field contributes to the DM candidate, which is a mixed, however mostly RH, sneutrino. 

The collider phenomenology of the MSSM+RN can be quite different from the typical MSSM case. It is therefore interesting and relevant to ask how the SUSY search results from Run~1 of the LHC, which were mostly designed with the MSSM in mind, constrain sneutrino LSP scenarios. 
To address this question, we used \smodels\ for testing the MSSM+RN against more than 60 results from CMS and ATLAS searches in the context of so-called Simplified Model Spectra (SMS). More precisely, by considering the model parameter space where the sneutrino is a good DM candidate compatible with all current constraints, we assessed 
1.)~the constraining power of the current SMS results on such scenarios and 
2.)~the most relevant signatures not covered by the SMS approach. 

Concerning point 1.), we found that the dilepton + MET searches are among the most relevant ones, constraining 
sneutrino masses up to about 210 GeV and mostly wino-like charginos up to $m_{\cha_1}\approx 440$ GeV. 
It is important to note here that this amounts to re-interpreting the ATLAS and CMS searches for 
$pp\to \tilde l^+\tilde l^-\to l^+ l^-\neu_1\neu_1$ in terms of $pp\to\chap_1\cham_1\to l^+ l^-\snu_l\snu_l$
(the validity of this is discussed in Appendix~\ref{sec:efficiency}). 
Hadronic SUSY searches  exclude gluinos masses up to $m_{\tilde{g}}\approx1200$ GeV and LSP masses up to $m_{\snu_1}\approx 500$~GeV. 
Nonetheless in general we find that only a very limited portion of the parameter space can be properly excluded by SMS results. For most points in the $(m_{\tilde{g}}, m_{\snu_1})$ or $(m_{\cha_1}, m_{\snu_1})$ planes there exist 
parameter combinations that allow to avoid all limits. 
Indeed, most of the parameter space is either allowed (SMS constraints exist for the specific topologies of the point 
but all $\sigma\times\br$ of these topologies are below their 95\% CL upper limits) or not tested at all (there are no existing SMS constraints for the specific topologies of the point or each topology has a $\sigma\times\br$ which is smaller than 1 event at LHC Run~1). 
Direct DM searches are complementary to the SMS constraints: many points that are not tested by SMS results can potentially be excluded by XENON1T. Vice versa, points well below the neutrino background, hence not reachable by future DM detectors, are already excluded by SMS results. 

The second main result of this paper concerns point 2.), \ie\ the study of the allowed points in terms of missing topologies. In the hadronic sector, pair-produced gluinos with masses well in the reach of LHC Run~1 are not constrained because 
they feature one or more of the following:
\begin{itemize}
\item additional leptons: since the gluino cannot directly decay into the sneutrino LSP, the hadronic final state is often  accompanied by leptons;
\item mixed topologies: each of the pair-produced gluinos undergoes a different decay;
\item the gluinos decay into $tb$ final states. 
\end{itemize}
None of these possibilities are covered by the current SMS results. Note here that the last two items are also common in the MSSM, as described in~\cite{Kraml:2013mwa}. 
For EW production, missing topologies include: 
\begin{itemize}
\item single leptons; 
\item single $W$s; 
\item different-flavour opposite-sign leptons; 
\item same-sign $W$'s or same-sign taus (accompanied respectively by additional leptons/taus, or $W$s).
\end{itemize} 
While such signatures have been searched for by the SUSY and/or exotics groups in ATLAS and CMS, 
the results do not exist in terms of appropriate SMS interpretations. Such an SMS interpretation would 
be very interesting in particular for the mono-lepton + MET case, which promises to have a considerable 
impact for constraining the MSSM+RN model.\footnote{This could be done analogous to the existing $\cha_1\neu_2$ ($\cha_1\to W^\pm\neu_1$, $\neu_2\to Z^0\neu_1$) simplified models that are already assessed by the ATLAS and CMS SUSY groups, but with the chargino decaying to 100\% into $l^\pm\snu_l$ and the neutralino decaying 100\% into $\nu\snu_l$. However, since the chargino and neutralino masses need not be degenerate, we propose to consider as a first step $\cha_1\neu_1$ production followed by $\cha_1\to l^\pm\snu_l$ and $\neu_1\to \nu\snu_l$. The cross section upper limits should be provided in the chargino- versus sneutrino mass plane for different neutralino masses, for the cases $l=e,\mu$ and $l=\tau$, and if computationally feasible also for $l=e,\mu,\tau$ assuming equal rates.} 

A final comment is in order. While the SMS approach is very convenient for the characterisation of new physics signatures and vast surveys of parameter spaces, it clearly has its limitations.   
Given the high interest in non-standard SUSY (and other new physics) scenarios, 
we urge the experimental collaborations to document their analyses in a way that they can conveniently 
be re-casted in public simulation frameworks like {\sc CheckMATE}~\cite{Drees:2013wra} or the 
{\sc MadAnalysis}\,5 PAD~\cite{Dumont:2014tja}. 
(See also the recommendations in  \cite{Dumont:2014tja} and \cite{Kraml:2012sg} in this context). 
This would allow to go beyond the limitations of SMS approach and give a much more rigorous assessment of the constraints in a large variety of new physics models, including the sneutrino DM scenario discussed in this paper.  
Unfortunately we are still a long way from this. 
%So far, the implementation and validation of ATLAS and CMS analyses for re-interpretation  in general contexts it is an extremely tedious task, because the information given in the experimental papers is often  incomplete. We can only hope that this will improve in the future. 

%%%%%%%%%%%%%%%%%%%%%%%%%%%%%%%%%%%%%%%%%%%%%%%%%%%%%%%%%%%%
\section*{Acknowledgements} 
%%%%%%%%%%%%%%%%%%%%%%%%%%%%%%%%%%%%%%%%%%%%%%%%%%%%%%%%%%%%

SK, SuK, and UL thank their colleagues from the {\sc SModelS} developer team, 
Andre Lessa, Veronica and Wolfgang Magerl, Michael Traub and Wolfgang Waltenberger for many helpful discussions. 
We also thank B.~Dumont, B. Fuks, E. Conte and D. Sengupta for helpful discussions on recasting 
with {\sc MadAnalysis}\,5. SuK moreover acknowledges discussions with R.~Schoeffbeck. 
 
This work was supported in part by the ANR project {\sc DMAstroLHC} and 
the ``Investissements d'avenir, Labex ENIGMASS''. 
CA is supported by the ERC project 267117 hosted by UPMC-Paris 6, PI J. Silk. 
MECC is supported by Funda\c{c}\~ao de Amparo \`a Pesquisa do Estado de S\~ao
Paulo (FAPESP).
SuK is supported by the ``New Frontiers'' program of the Austrian Academy of Sciences.
UL is supported by the Labex ENIGMASS; she also gratefully acknowledges the
hospitality of LPSC Grenoble during research visits prior to starting her PhD
thesis at the LPSC. 

\clearpage
%%%%%%%%%%%%%%%%%%%%%%%%%%%%%%%%%%%%%%%%%%%%%%%%%%%%%%%%%%%%
\appendix

\section{Validity of slepton search results for chargino-pair production with decay into lepton+sneutrino}\label{sec:efficiency}

In the spirit of the \smodels\ philosophy, we apply SMS constraints for the $\ell^+\ell^- + {\rm MET}$ topology,
obtained in the context of pair production of charged sleptons, $pp\to\tilde\ell^+\tilde\ell^-$ followed by 
$\tilde\ell^{\pm}\to \ell^\pm\tilde\chi^0_1$  to the case of chargino-pair production $pp\to\tilde\chi_1^+\tilde\chi_1^-$ followed by $\cha_1\to \ell^\pm\tilde\nu_{\ell1}$,
despite the opposite spin configuration.
This can only be valid if the signal selection efficiencies in both scenarios are comparable.
To test this assumption, we use the recast code~\cite{PAD-code-ATLAS-SUSY-2013-11} 
for ATLAS search in final states with two leptons and missing transverse momentum,  
ATLAS-SUSY-2013-11~\cite{Aad:2014vma}, which is available 
in the framework of the {\sc MadAnalysis}\,5 ``Public Analysis Database''~\cite{Dumont:2014tja}.   
We consider two benchmark scenarios in the simplified-model spirit, an MSSM one with  
$(m_{\tilde\ell^{\pm}},\ m_{\tilde\chi^0_1})=(270,\, 100)$~GeV and an MSSM+RN one with 
$(m_{\cha_1},\ m_{\tilde\nu_1})=(270,\, 100)$~GeV. 
Events are generated with {\sc MadGraph}~5~\cite{Alwall:2011uj,Alwall:2014hca} and 
{\sc Pythia}~6.4~\cite{Sjostrand:2006za} and then passed through {\sc Delphes}~3~\cite{deFavereau:2013fsa} 
for the simulation of the detector effects.\footnote{Note that for the reconstruction of events with a sneutrino LSP it is necessary to define the sneutrino as MET, by adding a corresponding {\tt EnergyFraction} entry in the {\sc Delphes} card.} 
For simplicity, in the following we restrict our study to pair-production of selectrons for the MSSM case, and 
pair-production of charginos decaying exclusively via electrons in the MSSM+RN case. 
 
\begin{figure}[h!]\centering
\includegraphics[width=0.5\textwidth]{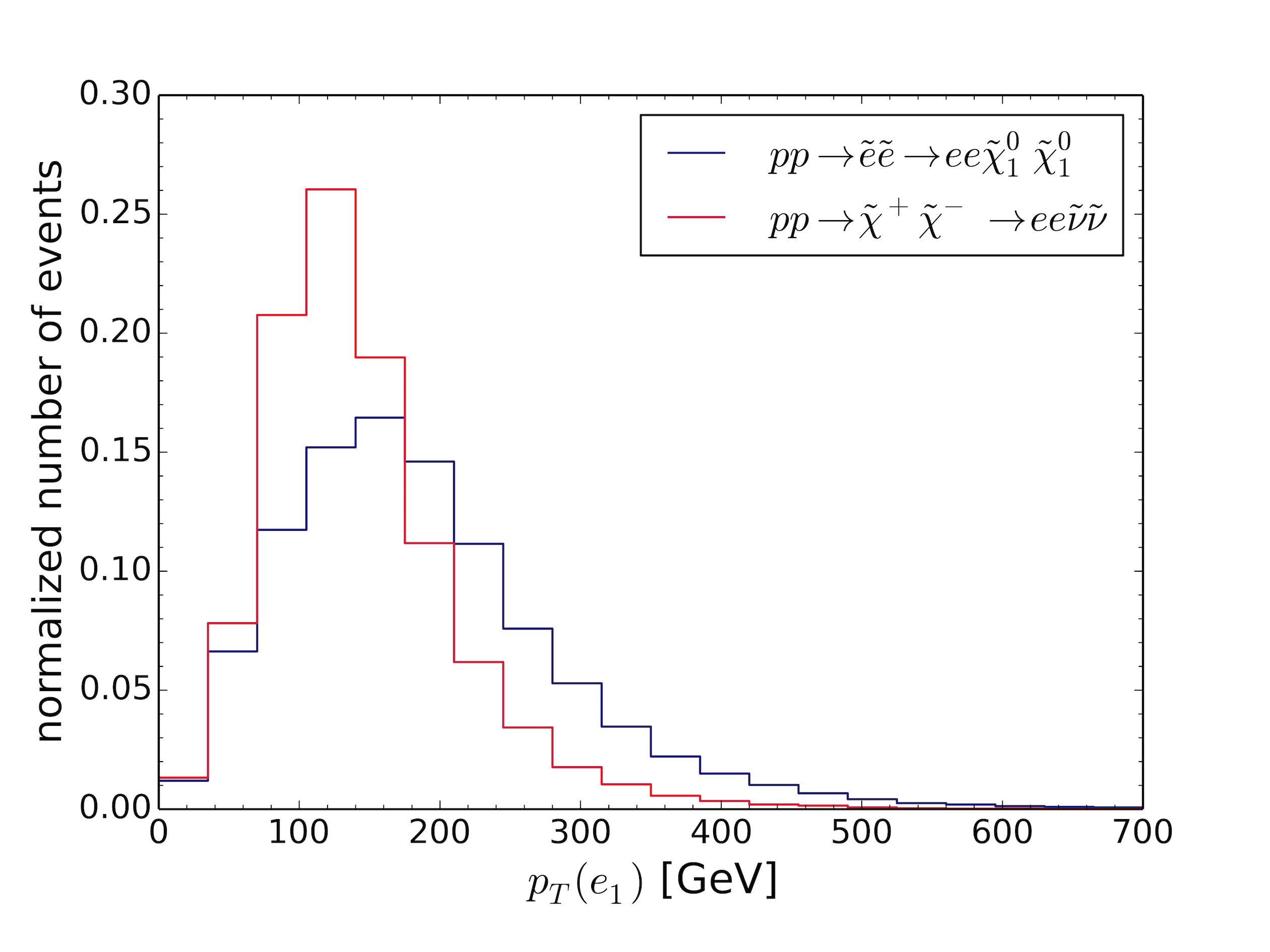}\hspace{-3mm}
\includegraphics[width=0.5\textwidth]{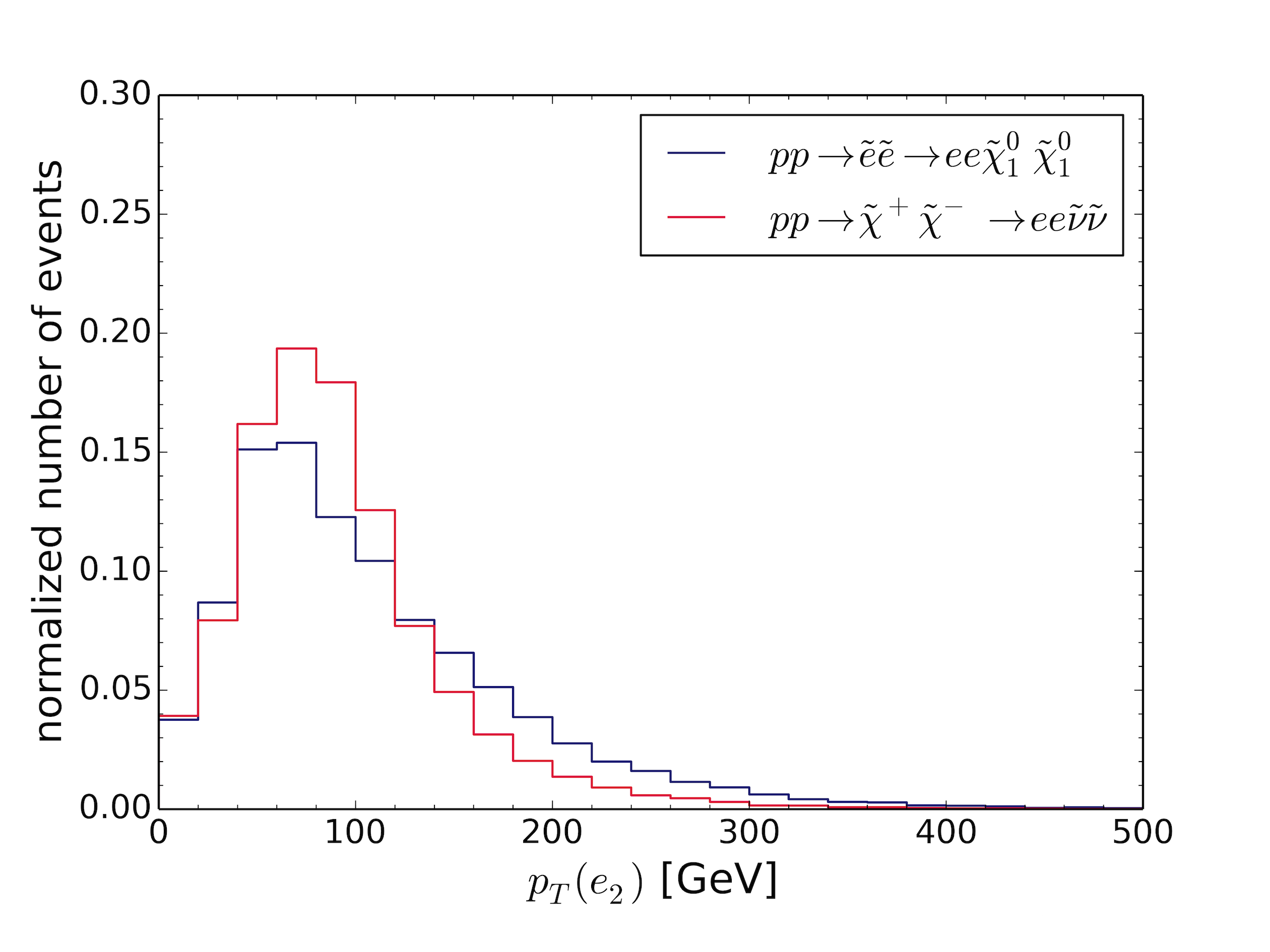} 
\caption{Comparison of the $p_T$ distributions of electrons originating from selectron decays in the 
MSSM and from chargino decays in MSSM+RN, at the level of reconstructed events. 
The benchmark scenarios used are $(m_{\tilde\ell^{\pm}},\ m_{\tilde\chi^0_1})=(270,\, 100)$~GeV for 
the MSSM case and $(m_{\cha_1},\ m_{\tilde\nu_1})=(270,\, 100)$~GeV for the MSSM+RN case. 
See text for details.}
\label{fig:leptonpT} 
\end{figure}

The event selection requires two opposite sign (OS), same flavor (SF) leptons with high transverse momentum, 
concretely $p_T > 35$~GeV and $p_T > 20$~GeV.\footnote{We consider here only the part of the analysis that 
is relevant for the SMS result used to constrain the sneutrino LSP scenario in Section~4.}   
Figure~\ref{fig:leptonpT} compares the $p_T$ distributions in the two benchmark scenarios, 
in the left panel for the harder electron, $e_1$, in the right panel for the second electron, $e_2$.
The bin sizes are chosen such that the first bin corresponds to the events that do not pass the 
$p_T > 35$~GeV (left panel) or $p_T > 20$~GeV (right panel) requirement. 
We see that the electrons originating from selectron-pair production tend to be harder than those 
originating from chargino-pair production. 

\begin{figure}[h!]\centering
\includegraphics[width=0.5\textwidth]{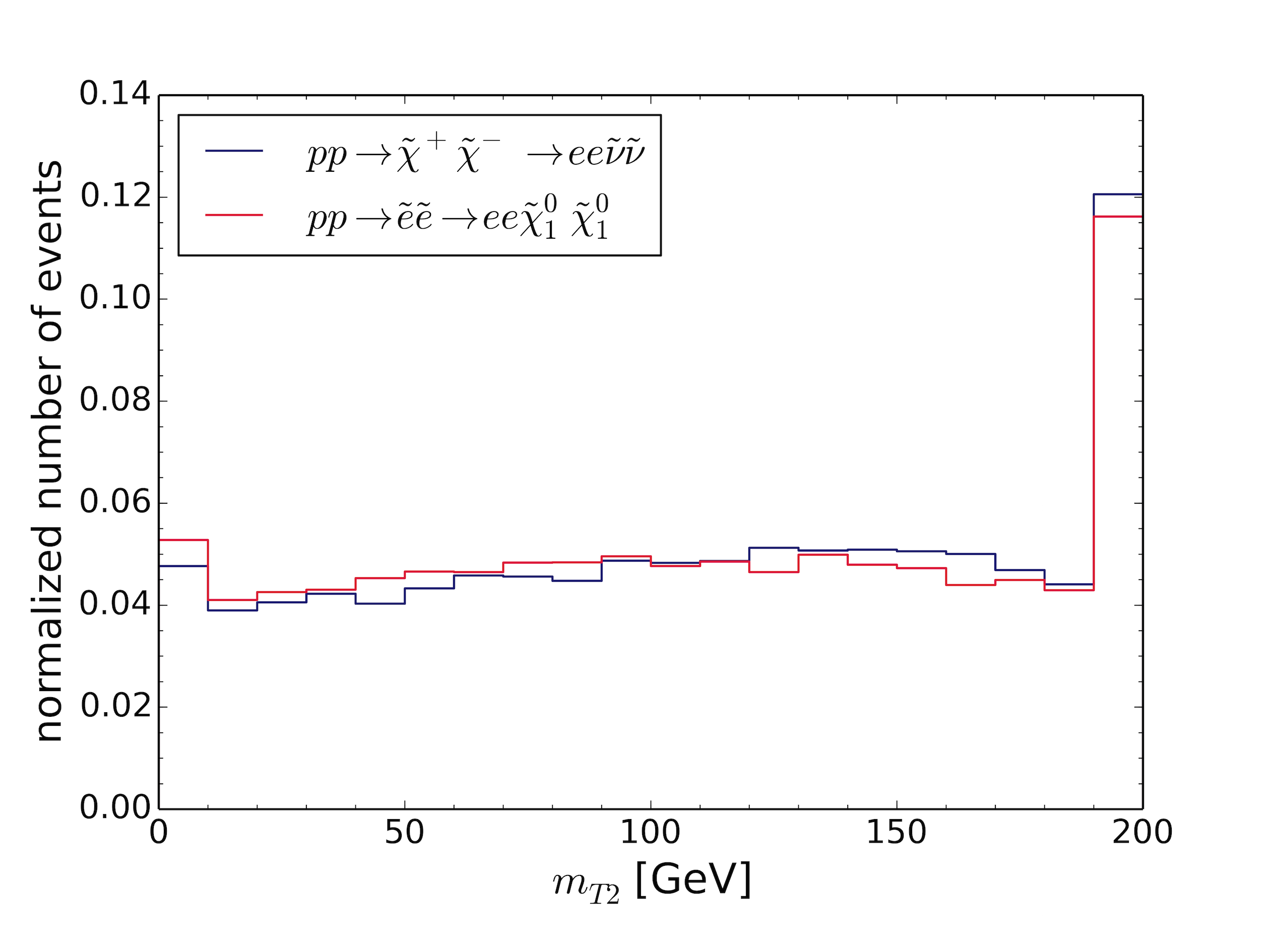}
\caption{Comparison of the $m_{T2}$ distributions for the two benchmark scenarios 
after all preselection cuts.}
\label{fig:mT2}
\end{figure}

The analysis further requires the invariant mass of the lepton pair to be outside the $Z$ window, 
and $\tau$s and jets are vetoed. Finally, three signal regions are defined by thresholds on the $m_{T2}$ 
(``stransverse mass'') variable~\cite{Lester:1999tx,Cheng:2008hk} that is used for reducing the $t\bar t$ and $Wt$ backgrounds: $m_{T2} > 90$, $> 120$ and $> 150$~GeV. 
The $m_{T2}$ distributions after the preselection cuts are shown in Fig.~\ref{fig:mT2}.
It can be seen that the distributions intersect around the minimum required value of $m_{T2} = 90$ GeV; 
events with electrons originating from chargino decays are more likely to pass this cut.

To see the net effect on the signal efficiencies, Table~\ref{tab:cutflow} shows the complete cut-flow comparison 
for the two benchmark scenarios. 
As expected, differences arise in the first cut, selecting high $p_T$ OS lepton pairs, and when applying 
the lower bounds for $m_{T2}$.
Because of the softer $p_T$ distribution in case of chargino production+decay, there are fewer events 
passing the first cut for this scenario. However, the opposite is true for the $m_{T2}$ cut.
Ultimately, the efficiencies are comparable in all signal regions, and even somewhat higher for the MSSM+RN scenario.

To check that this is still true closer to the kinematic edge, we reproduce the cut-flows for a second set 
of benchmark scenarios with an LSP mass of 200~GeV.  
As can be seen in Table~\ref{tab:cutflow2}, we find a similar behaviour in this case. 
We conclude that we can safely apply the SMS upper limits given by the experimental collaborations in the context of 
slepton-pair production in the MSSM to constrain chargino-pair production followed by decays into $l\snu_l$ 
in the MSSM+RN.

%\clearpage

\begin{table}[h!]
\caption{Comparison of the cut-flows for $pp\to\tilde e\tilde e\to e^+e^-\neu_1\neu_1$ and 
$pp\to\chap_1\cham_1\to e^+e^-\snu_{1}\snu_{1}$ with 
$(m_{\tilde\ell^{\pm}},\ m_{\tilde\chi^0_1})=(270,\, 100)$~GeV and 
$(m_{\cha_1},\ m_{\tilde\nu_1})=(270,\, 100)$~GeV, respectively.
\label{tab:cutflow}}
\begin{center} \renewcommand\arraystretch{1.2}
\begin{tabular}{| l | c | c  |}
  \hline
  Cut  & Slepton production & Chargino production \\
  \hline 
  \hline
\multicolumn{3}{|c|}{Common preselection}\\
 \hline
  Initial number of events & 50000 & 50000\\
  2 OS leptons & 35133 & 33464 \\
 $m_{ll} > 20$ GeV & 35038 & 33337 \\
 $\tau$ veto &  35007 & 33318 \\
 $ee$ leptons &  35007 & 33318 \\
 jet veto &  20176 & 19942   \\
 $Z$ veto &  19380  & 18984\\
\hline
\multicolumn{3}{|c|}{Different $m_{T2}$ regions}\\
\hline
 $m_{T2} > 90$ GeV & 11346  & 11594 \\
 $m_{T2} > 120$ GeV & 8520 & 8828  \\
 $m_{T2} > 150$ GeV & 5723 & 5926  \\
\hline
\end{tabular}
\end{center}\renewcommand\arraystretch{1.0}
\end{table}

\begin{table}[h!]
\caption{As Table~\ref{tab:cutflow} but for  
$(m_{\tilde\ell^{\pm}},\ m_{\tilde\chi^0_1})=(270,\, 200)$~GeV and  
$(m_{\cha_1},\ m_{\tilde\nu_1})=(270,\, 200)$~GeV. 
 \label{tab:cutflow2}}
\begin{center} \renewcommand\arraystretch{1.2}
\begin{tabular}{| l | c | c  |}
  \hline
  Cut  & Slepton production & Chargino production \\
  \hline
  \hline
\multicolumn{3}{|c|}{Common preselection}\\
 \hline
  Initial number of events & 50000 & 50000\\
  2 OS leptons & 29291 & 27244 \\
 $m_{ll} > 20$ GeV & 29082 & 26964 \\
 $\tau$ veto &  29050 & 26956 \\
 $ee$ leptons &  29050 & 26956 \\
 jet veto &  16834 & 16114  \\
 $Z$ veto &  15281  & 14025 \\
\hline
\multicolumn{3}{|c|}{Different $m_{T2}$ regions}\\
\hline
 $m_{T2} > 90$ GeV & 3028  & 3198 \\
 $m_{T2} > 120$ GeV & 85 & 140  \\
 $m_{T2} > 150$ GeV & 0 & 0  \\
\hline
\end{tabular}
\end{center}\renewcommand\arraystretch{1.0}
\end{table}

\clearpage
%%%%%%%%%%%%%%%%%%%%%%%%%%%%%%%%%%%%%%%%%%%%%%%%%%%%%%%%%%%%
\section{Lifetimes of long-lived particles}\label{sec:lifetime}

As mentioned in Sec.~\ref{sec:sms}, a considerable number of the scan points 
comprise long-lived sparticles. These occur mostly when enforcing light gluinos or squarks; 
in this case about 30~\% of the points feature long-lived particles,
while the fraction is below 1~\% without this constraint.
The long-lived particles are predominantly gluinos (85~\%),
mostly in the case where it is the NLSP, and in a few points where $\tilde{\chi}^0_1$ is slightly 
(up to about 50~GeV) lighter than the gluino.
Apart from that we find points with long-lived stops or staus in case they are the NLSP,
as well as single points with long-lived charginos.
Here we will focus on the long-lived gluinos and stops, long-lived staus have
been discussed before in \cite{Arina:2013zca}.\\

\begin{figure}[h]\centering
\includegraphics[width=0.52\textwidth]{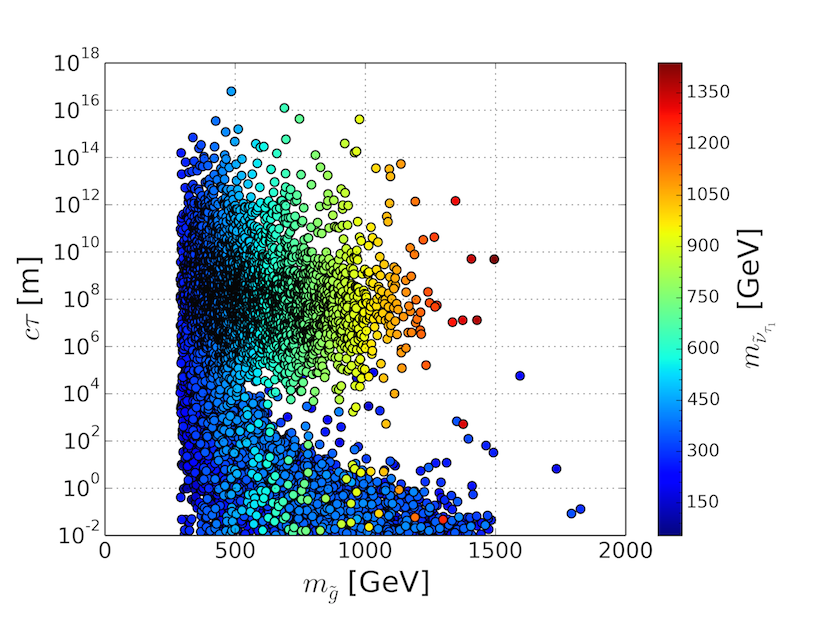}\hspace*{-5mm}
\includegraphics[width=0.52\textwidth]{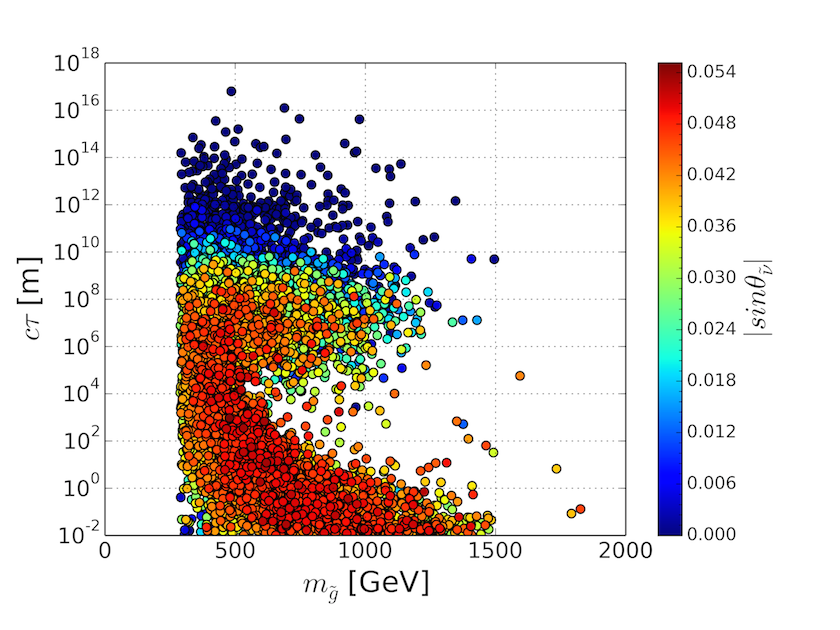}
\caption{Lifetimes $c\tau$ in [m] for long-lived gluinos, the color code indicates the LSP mass (left) and the sneutrino mixing angle (right).}
\label{fig:gluinoLT}
\end{figure}

In the MSSM  long-lived gluinos appear when all squarks are extremely heavy,
e.g. in split-SUSY scenarios.
In case of the MSSM+RN with a sneutrino LSP additional causes come into play.
If the gluino is the NLSP, its decay will proceed only via virtual squarks and gauginos,
yielding an effective four body decay, $\tilde{g} \rightarrow q q \nu \tilde{\nu}$ (virtual $\tilde{q}$ and $\tilde{\chi}^0$) or $\tilde{g} \rightarrow q q' l \tilde{\nu}$ (virtual $\tilde{q}$ and $\tilde{\chi}^{\pm}$).
The gluino lifetime will therefore depend not only on the squark mass, but also
on the gaugino masses and mixings, as well as the sneutrino mixing angle.
Meta-stable gluinos can thus appear even if the squarks are not completely decoupled.
The gluino lifetime as a function of its mass is shown in
Fig.~\ref{fig:gluinoLT}.
The left plot illustrates the depencence on the sneutrino mass, the right
plot the dependence on the sneutrino mixing.
We can distinguish two general regions.
First, we observe an exponential dependence of the lifetime on the gluino mass
for decay lengths of $10$~mm up to $10^4$ m.
Here the lifetime is largely independent of the sneutrino mass.
Moreover lifetimes at constant gluino masses are longer for
heavier squarks and gauginos.
In this region we generally find large mixing angles $\sin\theta_{\tilde{\nu}}$,
but heavy gauginos and squarks.
Points with very small mixing angles may also appear in this region,
in the case that the mass of the lightest neutralino is below the gluino mass.
The second region, with lifetimes longer than  $10^4$~m, and up to  $10^{17}$~m,
shows a very different behaviour.
We can see a clear correlation between gluino and sneutrino masses in this region,
with longer lifetimes found for smaller mass splittings.
The lifetimes moreover increase when going to very small sneutrino mixing
angles, with the maximum lifetimes achieved for $\sin\theta_{\tilde{\nu}}$ going to zero.

\begin{figure}[h]\centering
\includegraphics[width=0.52\textwidth]{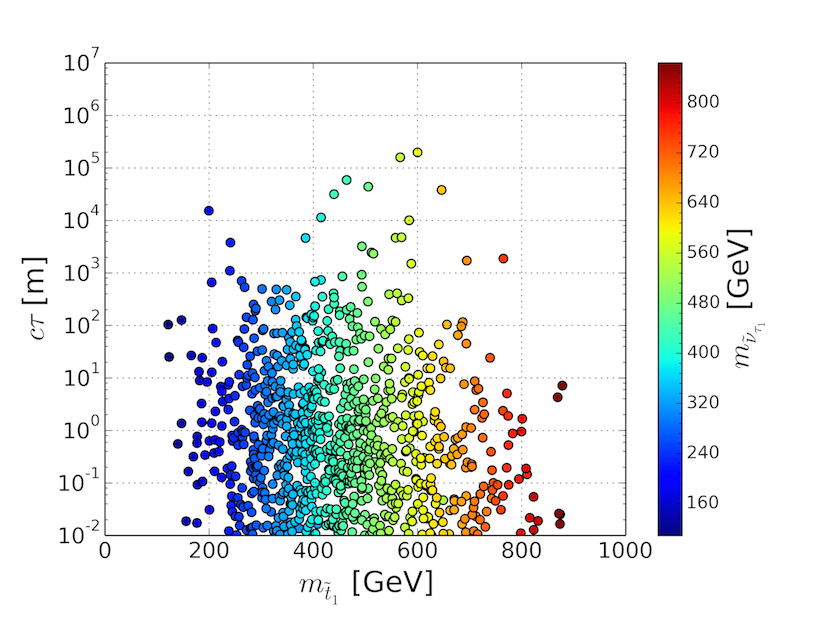}\hspace*{-5mm}
\includegraphics[width=0.52\textwidth]{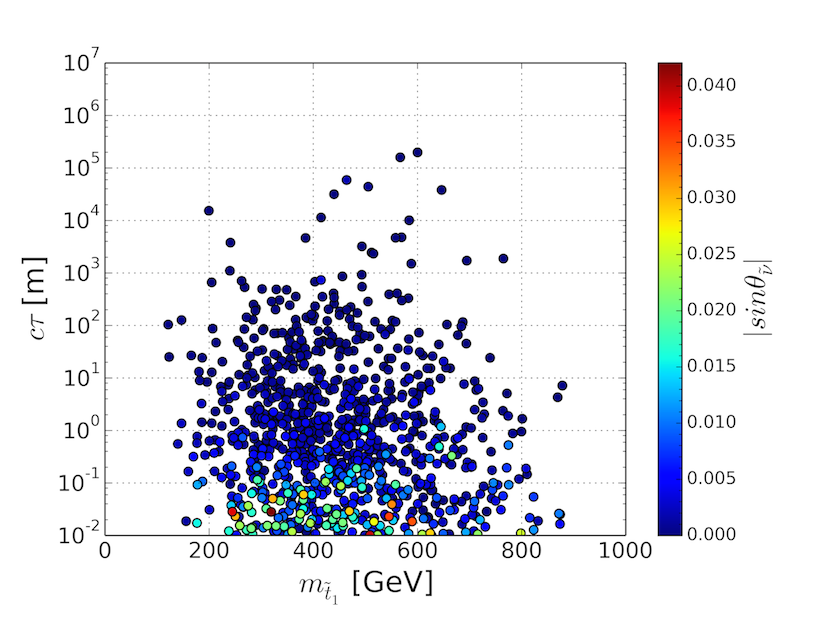}
\caption{Lifetimes $c\tau$ in [m] for long-lived stops, the color code indicates the LSP mass (left) and the sneutrino mixing angle (right).}
\label{fig:lifetime}
\end{figure}

Likewise, if the stop is the NLSP\footnote{If the stop mass is close to the gluino mass, both stop and gluino may be long-lived.} and has a small mass difference with the
sneutrino, it can be long-lived, see Fig.~\ref{fig:lifetime}.
As seen for the gluinos, the lifetime depends strongly on the sneutrino mixing.

Both long lived gluinos and long lived stops can be constrained by searches for R-hadrons, see~\cite{Chatrchyan:2013oca,ATLAS:2014fka} for R-hadrons escaping the detector, \cite{Aad:2013gva} for stopped R-hadrons,
or~\cite{ATLAS-CONF-2014-037} for metastable gluinos decaying in flight inside the
detector.
However, large uncertainties arise from modeling both the hadronisation
and the strong interaction of the R-hadron with the detector.
Therefore no collider constraints on long-lived sparticles have been included.\\
Additionally, cosmological constraints become important for gluino lifetimes
of about $100$~s ($10^{10}$~m)~\cite{Arvanitaki:2005fa}.
Lifetimes of that order would affect the fraction of heavy nuclei produced
during the Big Bang nucleosynthesis.
Longer lifetimes can further be constrained by searches for diffuse gamma ray
background, distortions in the CMBR and heavy isotopes.

\clearpage
%%%%%%%%%%%%%%%%%%%%%%%%%%%%%%%%%%%%%%%%
\bibliographystyle{JHEP}
\bibliography{biblio}

\providecommand{\href}[2]{#2}\begingroup\raggedright\begin{thebibliography}{10}

\bibitem{atlas:2012gk}
{\bf ATLAS} Collaboration, G.~Aad et~al., {\it {Observation of a new particle
  in the search for the Standard Model Higgs boson with the ATLAS detector at
  the LHC}},  {\em Phys.Lett.} {\bf B716} (2012) 1--29,
  [\href{http://xxx.lanl.gov/abs/1207.7214}{{\tt arXiv:1207.7214}}].

\bibitem{cms:2012gu}
{\bf CMS} Collaboration, S.~Chatrchyan et~al., {\it {Observation of a new boson
  at a mass of 125 GeV with the CMS experiment at the LHC}},  {\em Phys.Lett.}
  {\bf B716} (2012) 30--61, [\href{http://xxx.lanl.gov/abs/1207.7235}{{\tt
  arXiv:1207.7235}}].

\bibitem{atlas:susy:twiki}
https://twiki.cern.ch/twiki/bin/view/AtlasPublic/SupersymmetryPublicResults.

\bibitem{cms:susy:twiki}
https://twiki.cern.ch/twiki/bin/view/CMSPublic/PhysicsResultsSUS.

\bibitem{Djouadi:2013vqa}
A.~Djouadi and J.~Quevillon, {\it {The MSSM Higgs sector at a high $M_{SUSY}$:
  reopening the low tan$\beta$ regime and heavy Higgs searches}},  {\em JHEP}
  {\bf 1310} (2013) 028, [\href{http://xxx.lanl.gov/abs/1304.1787}{{\tt
  arXiv:1304.1787}}].

\bibitem{Giudice:2004tc}
G.~Giudice and A.~Romanino, {\it {Split supersymmetry}},  {\em Nucl.Phys.} {\bf
  B699} (2004) 65--89, [\href{http://xxx.lanl.gov/abs/hep-ph/0406088}{{\tt
  hep-ph/0406088}}].

\bibitem{Bhattacharyya:2012ct}
G.~Bhattacharyya and T.~S. Ray, {\it {Naturally split supersymmetry}},  {\em
  JHEP} {\bf 1205} (2012) 022, [\href{http://xxx.lanl.gov/abs/1201.1131}{{\tt
  arXiv:1201.1131}}].

\bibitem{Benakli:2013msa}
K.~Benakli, L.~DarmŽ, M.~D. Goodsell, and P.~Slavich, {\it {A Fake Split
  Supersymmetry Model for the 126 GeV Higgs}},  {\em JHEP} {\bf 1405} (2014)
  113, [\href{http://xxx.lanl.gov/abs/1312.5220}{{\tt arXiv:1312.5220}}].

\bibitem{Hall:2011jd}
L.~J. Hall and Y.~Nomura, {\it {Spread Supersymmetry}},  {\em JHEP} {\bf 1201}
  (2012) 082, [\href{http://xxx.lanl.gov/abs/1111.4519}{{\tt
  arXiv:1111.4519}}].

\bibitem{Dreiner:2012gx}
H.~K. Dreiner, M.~Kramer, and J.~Tattersall, {\it {How low can SUSY go?
  Matching, monojets and compressed spectra}},  {\em Europhys.Lett.} {\bf 99}
  (2012) 61001, [\href{http://xxx.lanl.gov/abs/1207.1613}{{\tt
  arXiv:1207.1613}}].

\bibitem{Fan:2011yu}
J.~Fan, M.~Reece, and J.~T. Ruderman, {\it {Stealth Supersymmetry}},  {\em
  JHEP} {\bf 1111} (2011) 012, [\href{http://xxx.lanl.gov/abs/1105.5135}{{\tt
  arXiv:1105.5135}}].

\bibitem{Borzumati:2000mc}
F.~Borzumati and Y.~Nomura, {\it {Low scale seesaw mechanisms for light
  neutrinos}},  {\em Phys.Rev.} {\bf D64} (2001) 053005,
  [\href{http://xxx.lanl.gov/abs/hep-ph/0007018}{{\tt hep-ph/0007018}}].

\bibitem{Arkani-Hamed:2000bq}
N.~Arkani-Hamed, L.~J. Hall, H.~Murayama, D.~R. Smith, and N.~Weiner, {\it
  Small neutrino masses from supersymmetry breaking},  {\em Phys. Rev.} {\bf
  D64} (2001) 115011, [\href{http://xxx.lanl.gov/abs/hep-ph/0006312}{{\tt
  hep-ph/0006312}}].

\bibitem{Hinshaw:2012aka}
{\bf WMAP} Collaboration, G.~Hinshaw et~al., {\it {Nine-Year Wilkinson
  Microwave Anisotropy Probe (WMAP) Observations: Cosmological Parameter
  Results}},  {\em Astrophys.J.Suppl.} {\bf 208} (2013) 19,
  [\href{http://xxx.lanl.gov/abs/1212.5226}{{\tt arXiv:1212.5226}}].

\bibitem{Ade:2013zuv}
{\bf Planck} Collaboration, P.~Ade et~al., {\it {Planck 2013 results. XVI.
  Cosmological parameters}},  {\em Astron.Astrophys.} (2014)
  [\href{http://xxx.lanl.gov/abs/1303.5076}{{\tt arXiv:1303.5076}}].

\bibitem{Asaka:2005cn}
T.~Asaka, K.~Ishiwata, and T.~Moroi, {\it Right-handed sneutrino as cold dark
  matter},  {\em Phys. Rev.} {\bf D73} (2006) 051301,
  [\href{http://xxx.lanl.gov/abs/hep-ph/0512118}{{\tt hep-ph/0512118}}].

\bibitem{Deppisch:2008bp}
F.~Deppisch and A.~Pilaftsis, {\it {Thermal Right-Handed Sneutrino Dark Matter
  in the F(D)-Term Model of Hybrid Inflation}},  {\em JHEP} {\bf 0810} (2008)
  080, [\href{http://xxx.lanl.gov/abs/0808.0490}{{\tt arXiv:0808.0490}}].

\bibitem{Cerdeno:2009dv}
D.~G. Cerdeno and O.~Seto, {\it {Right-handed sneutrino dark matter in the
  NMSSM}},  {\em JCAP} {\bf 0908} (2009) 032,
  [\href{http://xxx.lanl.gov/abs/0903.4677}{{\tt arXiv:0903.4677}}].

\bibitem{Khalil:2011tb}
S.~Khalil, H.~Okada, and T.~Toma, {\it {Right-handed Sneutrino Dark Matter in
  Supersymmetric B-L Model}},  {\em JHEP} {\bf 1107} (2011) 026,
  [\href{http://xxx.lanl.gov/abs/1102.4249}{{\tt arXiv:1102.4249}}].

\bibitem{Choi:2012ap}
K.-Y. Choi and O.~Seto, {\it {A Dirac right-handed sneutrino dark matter and
  its signature in the gamma-ray lines}},  {\em Phys.Rev.} {\bf D86} (2012)
  043515, [\href{http://xxx.lanl.gov/abs/1205.3276}{{\tt arXiv:1205.3276}}].

\bibitem{Belanger:2010cd}
G.~Belanger, M.~Kakizaki, E.~Park, S.~Kraml, and A.~Pukhov, {\it {Light mixed
  sneutrinos as thermal dark matter}},  {\em JCAP} {\bf 1011} (2010) 017,
  [\href{http://xxx.lanl.gov/abs/1008.0580}{{\tt arXiv:1008.0580}}].

\bibitem{Dumont:2012ee}
B.~Dumont, G.~Belanger, S.~Fichet, S.~Kraml, and T.~Schwetz, {\it {Mixed
  sneutrino dark matter in light of the 2011 XENON and LHC results}},  {\em
  JCAP} {\bf 1209} (2012) 013, [\href{http://xxx.lanl.gov/abs/1206.1521}{{\tt
  arXiv:1206.1521}}].

\bibitem{Hooper:2004dc}
D.~Hooper, J.~March-Russell, and S.~M. West, {\it Asymmetric sneutrino dark
  matter and the {O}mega(b)/{O}mega({DM}) puzzle},  {\em Phys. Lett.} {\bf
  B605} (2005) 228--236, [\href{http://xxx.lanl.gov/abs/hep-ph/0410114}{{\tt
  hep-ph/0410114}}].

\bibitem{Arina:2007tm}
C.~Arina and N.~Fornengo, {\it {Sneutrino cold dark matter, a new analysis:
  Relic abundance and detection rates}},  {\em JHEP} {\bf 0711} (2007) 029,
  [\href{http://xxx.lanl.gov/abs/0709.4477}{{\tt arXiv:0709.4477}}].

\bibitem{Thomas:2007bu}
Z.~Thomas, D.~Tucker-Smith, and N.~Weiner, {\it {Mixed Sneutrinos, Dark Matter
  and the CERN LHC}},  {\em Phys.Rev.} {\bf D77} (2008) 115015,
  [\href{http://xxx.lanl.gov/abs/0712.4146}{{\tt arXiv:0712.4146}}].

\bibitem{Belanger:2011ny}
G.~Belanger, S.~Kraml, and A.~Lessa, {\it {Light Sneutrino Dark Matter at the
  LHC}},  {\em JHEP} {\bf 1107} (2011) 083,
  [\href{http://xxx.lanl.gov/abs/1105.4878}{{\tt arXiv:1105.4878}}].

\bibitem{Arina:2013zca}
C.~Arina and M.~E. Cabrera, {\it {Multi-lepton signatures at LHC from sneutrino
  dark matter}},  {\em JHEP} {\bf 1404} (2014) 100,
  [\href{http://xxx.lanl.gov/abs/1311.6549}{{\tt arXiv:1311.6549}}].

\bibitem{BhupalDev:2012ru}
P.~Bhupal~Dev, S.~Mondal, B.~Mukhopadhyaya, and S.~Roy, {\it {Phenomenology of
  Light Sneutrino Dark Matter in cMSSM/mSUGRA with Inverse Seesaw}},  {\em
  JHEP} {\bf 1209} (2012) 110, [\href{http://xxx.lanl.gov/abs/1207.6542}{{\tt
  arXiv:1207.6542}}].

\bibitem{Guo:2013asa}
J.~Guo, Z.~Kang, J.~Li, T.~Li, and Y.~Liu, {\it {Simplified Supersymmetry with
  Sneutrino LSP at 8 TeV LHC}},  {\em JHEP} {\bf 1410} (2014) 164,
  [\href{http://xxx.lanl.gov/abs/1312.2821}{{\tt arXiv:1312.2821}}].

\bibitem{Harland-Lang:2013wxa}
L.~A. Harland-Lang, C.-H. Kom, K.~Sakurai, and M.~Tonini, {\it {Sharpening
  $m_{T2}$ cusps: the mass determination of semi-invisibly decaying particles
  from a resonance}},  {\em JHEP} {\bf 1406} (2014) 175,
  [\href{http://xxx.lanl.gov/abs/1312.5720}{{\tt arXiv:1312.5720}}].

\bibitem{Akerib:2013tjd}
{\bf LUX} Collaboration, D.~Akerib et~al., {\it {First results from the LUX
  dark matter experiment at the Sanford Underground Research Facility}},  {\em
  Phys.Rev.Lett.} {\bf 112} (2014) 091303,
  [\href{http://xxx.lanl.gov/abs/1310.8214}{{\tt arXiv:1310.8214}}].

\bibitem{Alwall:2008ag}
J.~Alwall, P.~Schuster, and N.~Toro, {\it {Simplified Models for a First
  Characterization of New Physics at the LHC}},  {\em Phys.Rev.} {\bf D79}
  (2009) 075020, [\href{http://xxx.lanl.gov/abs/0810.3921}{{\tt
  arXiv:0810.3921}}].

\bibitem{Alves:2011wf}
{\bf LHC New Physics Working Group} Collaboration, D.~Alves et~al., {\it
  {Simplified Models for LHC New Physics Searches}},  {\em J.Phys.} {\bf G39}
  (2012) 105005, [\href{http://xxx.lanl.gov/abs/1105.2838}{{\tt
  arXiv:1105.2838}}].

\bibitem{Kraml:2013mwa}
S.~Kraml, S.~Kulkarni, U.~Laa, A.~Lessa, W.~Magerl, D.~Proschofsky, and
  W.~Waltenberger, {\it {SModelS: a tool for interpreting simplified-model
  results from the LHC and its application to supersymmetry}},  {\em
  Eur.Phys.J.} {\bf C74} (2014) 2868,
  [\href{http://xxx.lanl.gov/abs/1312.4175}{{\tt arXiv:1312.4175}}].

\bibitem{smodels:v1}
S.~Kraml, S.~Kulkarni, U.~Laa, A.~Lessa, V.~Magerl, W.~Magerl, D.~Proschofsky,
  M.~Traub, and W.~Waltenberger, {\it {SModelS v1.0: a short user guide}},
  \href{http://xxx.lanl.gov/abs/1412.1745}{{\tt arXiv:1412.1745}}.

\bibitem{smodels:wiki}
{http://smodels.hephy.at}.

\bibitem{Smith:2001hy}
D.~R. Smith and N.~Weiner, {\it Inelastic dark matter},  {\em Phys. Rev.} {\bf
  D64} (2001) 043502, [\href{http://xxx.lanl.gov/abs/hep-ph/0101138}{{\tt
  hep-ph/0101138}}].

\bibitem{Grossman:1997is}
Y.~Grossman and H.~E. Haber, {\it Sneutrino mixing phenomena},  {\em Phys. Rev.
  Lett.} {\bf 78} (1997) 3438--3441,
  [\href{http://xxx.lanl.gov/abs/hep-ph/9702421}{{\tt hep-ph/9702421}}].

\bibitem{Aaij:2012nna}
{\bf LHCb} Collaboration, R.~Aaij et~al., {\it {First Evidence for the Decay
  $B^0_s \to \mu^+\mu^-$}},  {\em Phys.Rev.Lett.} {\bf 110} (2013) 021801,
  [\href{http://xxx.lanl.gov/abs/1211.2674}{{\tt arXiv:1211.2674}}].

\bibitem{Amhis:2012bh}
{\bf Heavy Flavor Averaging Group} Collaboration, Y.~Amhis et~al., {\it
  {Averages of B-Hadron, C-Hadron, and tau-lepton properties as of early
  2012}},  \href{http://xxx.lanl.gov/abs/1207.1158}{{\tt arXiv:1207.1158}}.

\bibitem{PDG}
{\bf Particle Data Group} Collaboration, K.~Olive et~al., {\it {Review of
  Particle Physics}},  {\em Chin.Phys.} {\bf C38} (2014) 090001.

\bibitem{Belanger:2013xza}
G.~Belanger, B.~Dumont, U.~Ellwanger, J.~Gunion, and S.~Kraml, {\it {Global fit
  to Higgs signal strengths and couplings and implications for extended Higgs
  sectors}},  {\em Phys.Rev.} {\bf D88} (2013) 075008,
  [\href{http://xxx.lanl.gov/abs/1306.2941}{{\tt arXiv:1306.2941}}].

\bibitem{sleptons2004}
{\bf {LEPSUSYWG, the ALEPH, DELPHI, L3 and OPAL Collaborations}} Collaboration.
\newblock LEPSUSYWG/04-01.1.

\bibitem{Abazov:2007aa}
{\bf D0} Collaboration, V.~Abazov et~al., {\it {Search for squarks and gluinos
  in events with jets and missing transverse energy using 2.1 $fb^{-1}$ of $p
  \bar{p}$ collision data at $\sqrt{s}$ = 1.96- TeV}},  {\em Phys.Lett.} {\bf
  B660} (2008) 449--457, [\href{http://xxx.lanl.gov/abs/0712.3805}{{\tt
  arXiv:0712.3805}}].

\bibitem{Cabrera:2012vu}
M.~E. Cabrera, J.~A. Casas, and R.~R. de~Austri, {\it {The health of SUSY after
  the Higgs discovery and the XENON100 data}},  {\em JHEP} {\bf 1307} (2013)
  182, [\href{http://xxx.lanl.gov/abs/1212.4821}{{\tt arXiv:1212.4821}}].

\bibitem{Allanach:2004rh}
B.~Allanach, A.~Djouadi, J.~Kneur, W.~Porod, and P.~Slavich, {\it {Precise
  determination of the neutral Higgs boson masses in the MSSM}},  {\em JHEP}
  {\bf 0409} (2004) 044, [\href{http://xxx.lanl.gov/abs/hep-ph/0406166}{{\tt
  hep-ph/0406166}}].

\bibitem{ALEPH:2005ab}
{\bf ALEPH Collaboration, DELPHI Collaboration, L3 Collaboration, OPAL
  Collaboration, SLD Collaboration, LEP Electroweak Working Group, SLD
  Electroweak Group, SLD Heavy Flavour Group} Collaboration, S.~Schael et~al.,
  {\it {Precision electroweak measurements on the $Z$ resonance}},  {\em
  Phys.Rept.} {\bf 427} (2006) 257--454,
  [\href{http://xxx.lanl.gov/abs/hep-ex/0509008}{{\tt hep-ex/0509008}}].

\bibitem{Boudjema:2011ig}
F.~Boudjema, G.~Drieu La~Rochelle, and S.~Kulkarni, {\it {One-loop corrections,
  uncertainties and approximations in neutralino annihilations: Examples}},
  {\em Phys.Rev.} {\bf D84} (2011) 116001,
  [\href{http://xxx.lanl.gov/abs/1108.4291}{{\tt arXiv:1108.4291}}].

\bibitem{Allanach:2001kg}
B.~Allanach, {\it {SOFTSUSY: a program for calculating supersymmetric
  spectra}},  {\em Comput.Phys.Commun.} {\bf 143} (2002) 305--331,
  [\href{http://xxx.lanl.gov/abs/hep-ph/0104145}{{\tt hep-ph/0104145}}].

\bibitem{Duhr:2011se}
C.~Duhr and B.~Fuks, {\it {A superspace module for the FeynRules package}},
  {\em Comput. Phys. Commun.} {\bf 182} (2011) 2404--2426,
  [\href{http://xxx.lanl.gov/abs/1102.4191}{{\tt arXiv:1102.4191}}].

\bibitem{Degrande:2011ua}
C.~Degrande, C.~Duhr, B.~Fuks, D.~Grellscheid, O.~Mattelaer, et~al., {\it {UFO
  - The Universal FeynRules Output}},  {\em Comput.Phys.Commun.} {\bf 183}
  (2012) 1201--1214, [\href{http://xxx.lanl.gov/abs/1108.2040}{{\tt
  arXiv:1108.2040}}].

\bibitem{Belanger:2013oya}
G.~Belanger, F.~Boudjema, A.~Pukhov, and A.~Semenov, {\it {micrOMEGAs\,3: A
  program for calculating dark matter observables}},  {\em Comput.Phys.Commun.}
  {\bf 185} (2014) 960--985, [\href{http://xxx.lanl.gov/abs/1305.0237}{{\tt
  arXiv:1305.0237}}].

\bibitem{Mahmoudi:2008tp}
F.~Mahmoudi, {\it {SuperIso v2.3: A Program for calculating flavor physics
  observables in Supersymmetry}},  {\em Comput.Phys.Commun.} {\bf 180} (2009)
  1579--1613, [\href{http://xxx.lanl.gov/abs/0808.3144}{{\tt
  arXiv:0808.3144}}].

\bibitem{Feroz:2007kg}
F.~Feroz and M.~Hobson, {\it {Multimodal nested sampling: an efficient and
  robust alternative to MCMC methods for astronomical data analysis}},  {\em
  Mon.Not.Roy.Astron.Soc.} {\bf 384} (2008) 449,
  [\href{http://xxx.lanl.gov/abs/0704.3704}{{\tt arXiv:0704.3704}}].

\bibitem{Feroz:2008xx}
F.~Feroz, M.~Hobson, and M.~Bridges, {\it {MultiNest: an efficient and robust
  Bayesian inference tool for cosmology and particle physics}},  {\em
  Mon.Not.Roy.Astron.Soc.} {\bf 398} (2009) 1601--1614,
  [\href{http://xxx.lanl.gov/abs/0809.3437}{{\tt arXiv:0809.3437}}].

\bibitem{2013arXiv1306.2144F}
F.~{Feroz}, M.~P. {Hobson}, E.~{Cameron}, and A.~N. {Pettitt}, {\it {Importance
  Nested Sampling and the MultiNest Algorithm}},  {\em ArXiv e-prints} (June,
  2013) [\href{http://xxx.lanl.gov/abs/1306.2144}{{\tt arXiv:1306.2144}}].

\bibitem{Sjostrand:2006za}
T.~Sjostrand, S.~Mrenna, and P.~Z. Skands, {\it {PYTHIA 6.4 Physics and
  Manual}},  {\em JHEP} {\bf 0605} (2006) 026,
  [\href{http://xxx.lanl.gov/abs/hep-ph/0603175}{{\tt hep-ph/0603175}}].

\bibitem{Beenakker:1996ch}
W.~Beenakker, R.~Hopker, M.~Spira, and P.~Zerwas, {\it {Squark and gluino
  production at hadron colliders}},  {\em Nucl.Phys.} {\bf B492} (1997)
  51--103, [\href{http://xxx.lanl.gov/abs/hep-ph/9610490}{{\tt
  hep-ph/9610490}}].

\bibitem{Beenakker:1997ut}
W.~Beenakker, M.~Kramer, T.~Plehn, M.~Spira, and P.~Zerwas, {\it {Stop
  production at hadron colliders}},  {\em Nucl.Phys.} {\bf B515} (1998) 3--14,
  [\href{http://xxx.lanl.gov/abs/hep-ph/9710451}{{\tt hep-ph/9710451}}].

\bibitem{Kulesza:2008jb}
A.~Kulesza and L.~Motyka, {\it {Threshold resummation for squark-antisquark and
  gluino-pair production at the LHC}},  {\em Phys.Rev.Lett.} {\bf 102} (2009)
  111802, [\href{http://xxx.lanl.gov/abs/0807.2405}{{\tt arXiv:0807.2405}}].

\bibitem{Kulesza:2009kq}
A.~Kulesza and L.~Motyka, {\it {Soft gluon resummation for the production of
  gluino-gluino and squark-antisquark pairs at the LHC}},  {\em Phys.Rev.} {\bf
  D80} (2009) 095004, [\href{http://xxx.lanl.gov/abs/0905.4749}{{\tt
  arXiv:0905.4749}}].

\bibitem{Beenakker:2009ha}
W.~Beenakker, S.~Brensing, M.~Kramer, A.~Kulesza, E.~Laenen, et~al., {\it
  {Soft-gluon resummation for squark and gluino hadroproduction}},  {\em JHEP}
  {\bf 0912} (2009) 041, [\href{http://xxx.lanl.gov/abs/0909.4418}{{\tt
  arXiv:0909.4418}}].

\bibitem{Beenakker:2010nq}
W.~Beenakker, S.~Brensing, M.~Kramer, A.~Kulesza, E.~Laenen, et~al., {\it
  {Supersymmetric top and bottom squark production at hadron colliders}},  {\em
  JHEP} {\bf 1008} (2010) 098, [\href{http://xxx.lanl.gov/abs/1006.4771}{{\tt
  arXiv:1006.4771}}].

\bibitem{Beenakker:2011fu}
W.~Beenakker, S.~Brensing, M.~Kramer, A.~Kulesza, E.~Laenen, et~al., {\it
  {Squark and Gluino Hadroproduction}},  {\em Int.J.Mod.Phys.} {\bf A26} (2011)
  2637--2664, [\href{http://xxx.lanl.gov/abs/1105.1110}{{\tt
  arXiv:1105.1110}}].

\bibitem{Buckley:2013jua}
A.~Buckley, {\it {PySLHA: a Pythonic interface to SUSY Les Houches Accord
  data}},  \href{http://xxx.lanl.gov/abs/1305.4194}{{\tt arXiv:1305.4194}}.

\bibitem{Skands:2003cj}
P.~Z. Skands, B.~Allanach, H.~Baer, C.~Balazs, G.~Belanger, et~al., {\it {SUSY
  Les Houches accord: Interfacing SUSY spectrum calculators, decay packages,
  and event generators}},  {\em JHEP} {\bf 0407} (2004) 036,
  [\href{http://xxx.lanl.gov/abs/hep-ph/0311123}{{\tt hep-ph/0311123}}].

\bibitem{slha-xsext}
``{Extending the SLHA: cross section information}.''
\newblock {http://phystev.in2p3.fr/wiki/2013:groups:tools:slha}.

\bibitem{Aad:2013wta}
{\bf ATLAS} Collaboration, G.~Aad et~al., {\it {Search for new phenomena in
  final states with large jet multiplicities and missing transverse momentum at
  $\sqrt{s}$=8 TeV proton-proton collisions using the ATLAS experiment}},  {\em
  JHEP} {\bf 1310} (2013) 130, [\href{http://xxx.lanl.gov/abs/1308.1841}{{\tt
  arXiv:1308.1841}}].

\bibitem{Aad:2014wea}
{\bf ATLAS} Collaboration, G.~Aad et~al., {\it {Search for squarks and gluinos
  with the ATLAS detector in final states with jets and missing transverse
  momentum using $\sqrt{s}=8$ TeV proton--proton collision data}},  {\em JHEP}
  {\bf 1409} (2014) 176, [\href{http://xxx.lanl.gov/abs/1405.7875}{{\tt
  arXiv:1405.7875}}].

\bibitem{ATLAS-CONF-2013-061}
{\bf ATLAS} Collaboration, {\it {Search for strong production of supersymmetric
  particles in final states with missing transverse momentum and at least three
  b-jets using 20.1~fb$^{-1}$ of pp collisions at sqrt(s) = 8 TeV with the
  ATLAS Detector.}},  Tech. Rep. ATLAS-CONF-2013-061, CERN, Geneva, Jun, 2013.

\bibitem{Chatrchyan:2013wxa}
{\bf CMS} Collaboration, S.~Chatrchyan et~al., {\it {Search for gluino mediated
  bottom- and top-squark production in multijet final states in pp collisions
  at 8 TeV}},  {\em Phys.Lett.} {\bf B725} (2013) 243--270,
  [\href{http://xxx.lanl.gov/abs/1305.2390}{{\tt arXiv:1305.2390}}].

\bibitem{Chatrchyan:2013lya}
{\bf CMS} Collaboration, S.~Chatrchyan et~al., {\it {Search for supersymmetry
  in hadronic final states with missing transverse energy using the variables
  $\alpha_T$ and b-quark multiplicity in pp collisions at $\sqrt s=8$ TeV}},
  {\em Eur.Phys.J.} {\bf C73} (2013), no.~9 2568,
  [\href{http://xxx.lanl.gov/abs/1303.2985}{{\tt arXiv:1303.2985}}].

\bibitem{Chatrchyan:2013iqa}
{\bf CMS} Collaboration, S.~Chatrchyan et~al., {\it {Search for supersymmetry
  in pp collisions at $\sqrt{s}$=8 TeV in events with a single lepton, large
  jet multiplicity, and multiple b jets}},  {\em Phys.Lett.} {\bf B733} (2014)
  328--353, [\href{http://xxx.lanl.gov/abs/1311.4937}{{\tt arXiv:1311.4937}}].

\bibitem{Chatrchyan:2014lfa}
{\bf CMS} Collaboration, S.~Chatrchyan et~al., {\it {Search for new physics in
  the multijet and missing transverse momentum final state in proton-proton
  collisions at $\sqrt{s}$= 8 TeV}},  {\em JHEP} {\bf 1406} (2014) 055,
  [\href{http://xxx.lanl.gov/abs/1402.4770}{{\tt arXiv:1402.4770}}].

\bibitem{Aad:2014vma}
{\bf ATLAS} Collaboration, G.~Aad et~al., {\it {Search for direct production of
  charginos, neutralinos and sleptons in final states with two leptons and
  missing transverse momentum in $pp$ collisions at $\sqrt{s} =$ 8 TeV with the
  ATLAS detector}},  {\em JHEP} {\bf 1405} (2014) 071,
  [\href{http://xxx.lanl.gov/abs/1403.5294}{{\tt arXiv:1403.5294}}].

\bibitem{Khachatryan:2014qwa}
{\bf CMS} Collaboration, V.~Khachatryan et~al., {\it {Searches for electroweak
  production of charginos, neutralinos, and sleptons decaying to leptons and W,
  Z, and Higgs bosons in pp collisions at 8 TeV}},  {\em Eur.Phys.J.} {\bf C74}
  (2014), no.~9 3036, [\href{http://xxx.lanl.gov/abs/1405.7570}{{\tt
  arXiv:1405.7570}}].

\bibitem{Aad:2014yka}
{\bf ATLAS} Collaboration, G.~Aad et~al., {\it {Search for the direct
  production of charginos, neutralinos and staus in final states with at least
  two hadronically decaying taus and missing transverse momentum in $pp$
  collisions at $\sqrt{s}$ = 8 TeV with the ATLAS detector}},  {\em JHEP} {\bf
  1410} (2014) 96, [\href{http://xxx.lanl.gov/abs/1407.0350}{{\tt
  arXiv:1407.0350}}].

\bibitem{ATLAS:2014wra}
{\bf ATLAS} Collaboration, G.~Aad et~al., {\it {Search for new particles in
  events with one lepton and missing transverse momentum in $pp$ collisions at
  $\sqrt{s}$ = 8 TeV with the ATLAS detector}},  {\em JHEP} {\bf 1409} (2014)
  037, [\href{http://xxx.lanl.gov/abs/1407.7494}{{\tt arXiv:1407.7494}}].

\bibitem{Khachatryan:2014tva}
{\bf CMS} Collaboration, V.~Khachatryan et~al., {\it {Search for physics beyond
  the standard model in final states with a lepton and missing transverse
  energy in proton-proton collisions at $\sqrt{s}$ = 8 TeV}},
  \href{http://xxx.lanl.gov/abs/1408.2745}{{\tt arXiv:1408.2745}}.

\bibitem{Aprile:2012zx}
{\bf XENON1T} Collaboration, E.~Aprile, {\it {The XENON1T Dark Matter Search
  Experiment}},  \href{http://xxx.lanl.gov/abs/1206.6288}{{\tt
  arXiv:1206.6288}}.

\bibitem{Billard:2013qya}
J.~Billard, L.~Strigari, and E.~Figueroa-Feliciano, {\it {Implication of
  neutrino backgrounds on the reach of next generation dark matter direct
  detection experiments}},  {\em Phys.Rev.} {\bf D89} (2014) 023524,
  [\href{http://xxx.lanl.gov/abs/1307.5458}{{\tt arXiv:1307.5458}}].

\bibitem{Drees:2013wra}
M.~Drees, H.~Dreiner, D.~Schmeier, J.~Tattersall, and J.~S. Kim, {\it
  {CheckMATE: Confronting your Favourite New Physics Model with LHC Data}},
  {\em Comput.Phys.Commun.} {\bf 187} (2014) 227--265,
  [\href{http://xxx.lanl.gov/abs/1312.2591}{{\tt arXiv:1312.2591}}].

\bibitem{Dumont:2014tja}
B.~Dumont, B.~Fuks, S.~Kraml, S.~Bein, G.~Chalons, et~al., {\it {Toward a
  public analysis database for LHC new physics searches using MADANALYSIS 5}},
  {\em Eur.Phys.J.} {\bf C75} (2015), no.~2 56,
  [\href{http://xxx.lanl.gov/abs/1407.3278}{{\tt arXiv:1407.3278}}].

\bibitem{Kraml:2012sg}
S.~Kraml, B.~Allanach, M.~Mangano, H.~Prosper, S.~Sekmen, et~al., {\it
  {Searches for New Physics: Les Houches Recommendations for the Presentation
  of LHC Results}},  {\em Eur.Phys.J.} {\bf C72} (2012) 1976,
  [\href{http://xxx.lanl.gov/abs/1203.2489}{{\tt arXiv:1203.2489}}].

\bibitem{PAD-code-ATLAS-SUSY-2013-11}
B.~Dumont, ``{MadAnalysis 5 implementation of ATLAS-SUSY-2013-11: di-leptons
  plus MET}.''
\newblock {DOI: 10.7484/INSPIREHEP.DATA.HLMR.T56W.2}.

\bibitem{Alwall:2011uj}
J.~Alwall, M.~Herquet, F.~Maltoni, O.~Mattelaer, and T.~Stelzer, {\it {MadGraph
  5 : Going Beyond}},  {\em JHEP} {\bf 1106} (2011) 128,
  [\href{http://xxx.lanl.gov/abs/1106.0522}{{\tt arXiv:1106.0522}}].

\bibitem{Alwall:2014hca}
J.~Alwall, R.~Frederix, S.~Frixione, V.~Hirschi, F.~Maltoni, et~al., {\it {The
  automated computation of tree-level and next-to-leading order differential
  cross sections, and their matching to parton shower simulations}},
  \href{http://xxx.lanl.gov/abs/1405.0301}{{\tt arXiv:1405.0301}}.

\bibitem{deFavereau:2013fsa}
J.~de~Favereau, C.~Delaere, P.~Demin, A.~Giammanco, V.~Lema�tre, et~al., {\it
  {DELPHES 3, A modular framework for fast simulation of a generic collider
  experiment}},  \href{http://xxx.lanl.gov/abs/1307.6346}{{\tt
  arXiv:1307.6346}}.

\bibitem{Lester:1999tx}
C.~Lester and D.~Summers, {\it {Measuring masses of semiinvisibly decaying
  particles pair produced at hadron colliders}},  {\em Phys.Lett.} {\bf B463}
  (1999) 99--103, [\href{http://xxx.lanl.gov/abs/hep-ph/9906349}{{\tt
  hep-ph/9906349}}].

\bibitem{Cheng:2008hk}
H.-C. Cheng and Z.~Han, {\it {Minimal Kinematic Constraints and $m_{T2}$}},
  {\em JHEP} {\bf 0812} (2008) 063,
  [\href{http://xxx.lanl.gov/abs/0810.5178}{{\tt arXiv:0810.5178}}].

\bibitem{Chatrchyan:2013oca}
{\bf CMS} Collaboration, S.~Chatrchyan et~al., {\it {Searches for long-lived
  charged particles in pp collisions at $\sqrt{s}$=7 and 8 TeV}},  {\em JHEP}
  {\bf 1307} (2013) 122, [\href{http://xxx.lanl.gov/abs/1305.0491}{{\tt
  arXiv:1305.0491}}].

\bibitem{ATLAS:2014fka}
{\bf ATLAS} Collaboration, G.~Aad et~al., {\it {Searches for heavy long-lived
  charged particles with the ATLAS detector in proton-proton collisions at $
  \sqrt{s}=8 $ TeV}},  {\em JHEP} {\bf 1501} (2015) 068,
  [\href{http://xxx.lanl.gov/abs/1411.6795}{{\tt arXiv:1411.6795}}].

\bibitem{Aad:2013gva}
{\bf ATLAS} Collaboration, G.~Aad et~al., {\it {Search for long-lived stopped
  R-hadrons decaying out-of-time with pp collisions using the ATLAS detector}},
   {\em Phys.Rev.} {\bf D88} (2013), no.~11 112003,
  [\href{http://xxx.lanl.gov/abs/1310.6584}{{\tt arXiv:1310.6584}}].

\bibitem{ATLAS-CONF-2014-037}
{\bf ATLAS} Collaboration, {\it {Limits on metastable gluinos from ATLAS SUSY
  searches at 8 TeV}},  Tech. Rep. ATLAS-CONF-2014-037, CERN, Geneva, Jul,
  2014.

\bibitem{Arvanitaki:2005fa}
A.~Arvanitaki, C.~Davis, P.~W. Graham, A.~Pierce, and J.~G. Wacker, {\it
  {Limits on split supersymmetry from gluino cosmology}},  {\em Phys.Rev.} {\bf
  D72} (2005) 075011, [\href{http://xxx.lanl.gov/abs/hep-ph/0504210}{{\tt
  hep-ph/0504210}}].

\end{thebibliography}\endgroup

%%%%%%%%%%%%%%%%%%%%%%%%%%%%%%%%%%%%%%%%

\end{document}